\pdfoutput=1

\documentclass[draft]{agujournal2019}
\usepackage{url} 
\usepackage{lineno}
\usepackage[inline]{trackchanges} 
\usepackage{soul}
\usepackage{amsmath,booktabs,hanging,array}
\draftfalse

\journalname{arXiv.org}

\begin{document}

%
%


\title{Further analysis of cGAN: A system for Generative Deep Learning Post-processing of Precipitation}

%
%




\authors{
Fenwick C. Cooper\affil{1}, Andrew T. T. McRae\affil{1}, Matthew Chantry\affil{2}, Bobby Antonio\affil{3} and Tim N.
Palmer\affil{1}
}


\affiliation{1}{Department of Physics, University of Oxford, Oxford, UK}
\affiliation{2}{European Centre for Medium-Range Weather Forecasts, Reading, UK}
\affiliation{3}{School of Geographical Sciences, University of Bristol, UK}




\correspondingauthor{F. C. Cooper}{fenwick.cooper@physics.ox.ac.uk}



\begin{keypoints}
\item The performance of a rainfall post-processing model (cGAN) is examined at three locations across the North America and compared to the UK.
\item cGAN trained on local data was competitive with the IFS ensemble forecast. cGAN trained on all regions had the best performance.
\item Training on ECMWF IFS forecasts leads to a CRPS that is lower (better) than training on ERA5.
\end{keypoints}

%
%

%
%


\begin{abstract}
The conditional generative adversarial rainfall model ``cGAN'' developed for the UK \cite{Harris22} was trained to post-process into an ensemble and downscale ERA5 rainfall to 1km resolution over three regions of the USA and the UK. Relative to radar data (stage IV and NIMROD), the quality of the forecast rainfall distribution was quantified locally at each grid point and between grid points using the spatial correlation structure. Despite only having information from a single lower quality analysis, the ensembles of post processed rainfall produced were found to be competitive with IFS ensemble forecasts with lead times of between 8 and 16 hours. Comparison to the original cGAN trained on the UK using the IFS HRES forecast indicates that improved training forecasts result in improved post-processing.

The cGAN models were additionally applied to the regions that they were not trained on. Each model performed well in their own region indicating that each model is somewhat region specific. However the model trained on the Washington DC, Atlantic coast, region achieved good scores across the USA and was competitive over the UK. There are more overall rainfall events spread over the whole region so the improved scores might be simply due to increased data. A model was therefore trained using data from all four regions which then outperformed the models trained locally.
\end{abstract}


%
%

%


%
%
%
%

\section{Introduction}

Given the measurements we have it is impossible to perfectly predict the exact amount of rainfall at some time in the future. Instead we try to predict a distribution of rainfall with ensemble methods. Generative-adversarial networks or GANs \cite{Goodfellow14} have been introduced as a method for approximating distributions and have been further developed to incorporate conditioning information \cite{Mirza14}. \citeA{Leinonen21} conditioned the empirical rainfall distribution on smoothed rainfall radar data using a GAN to ``downscale'' and reproduce the original un-smoothed data. This work was extended by \citeA{Harris22} to downscale and post-process ECMWF forecast data towards radar data over the UK. Similar work was performed independently by \citeA{Price22} at the same time. For a more in depth review of the background literature see \citeA{Harris22}.

Since then \citeA{Yang23} applied a GAN to post-process precipitation forecasts over China and \citeA{leinonen23} developed a diffusion model for precipitation nowcasting which shows some advantages in comparison to the equivalent GAN models. Another neural network based approach has been developed to post-process global medium range forecasts of precipitable water \cite{Agrawal23}. A large neural network model has been developed over the USA \cite{Andrychowicz23} that rather than post-processing computes the entire forecast of precipitation up to 24 hours ahead and a similar model has been favourably assessed in an operational setting \cite{BenBouallegue23}.

The goal of this paper is to further test the model we call cGAN developed in \citeA{Harris22} to find out where it does well and where it can be improved. In addition to the ECMWF IFS HRES deterministic forecast, the output of the cGAN, trained to correct ERA5 data with respect to higher resolution rainfall radar data, is compared to the 6-18 hour ECMWF IFS ensemble forecast predictions. In addition to looking at the UK, we add three regions of the USA and compare models trained on them separately and look at how good these models are outside of their training region. We focus on four metrics for different aspects of the quality of the produced rainfall distribution; the Cumulative Rank Probability Score (CRPS) and Root-Mean-Squared Error of the Ensemble Mean (RMSEEM) which measure the quality of the one dimensional conditional distribution at individual points, and the Radially Averaged Log Spectral Distance (RALSD) and Variogram score, both of which measure the quality of the spatial relationship between rainfall points.

\section{cGAN USA}

\begin{figure}
\begin{center}
\includegraphics[width=0.89\textwidth]{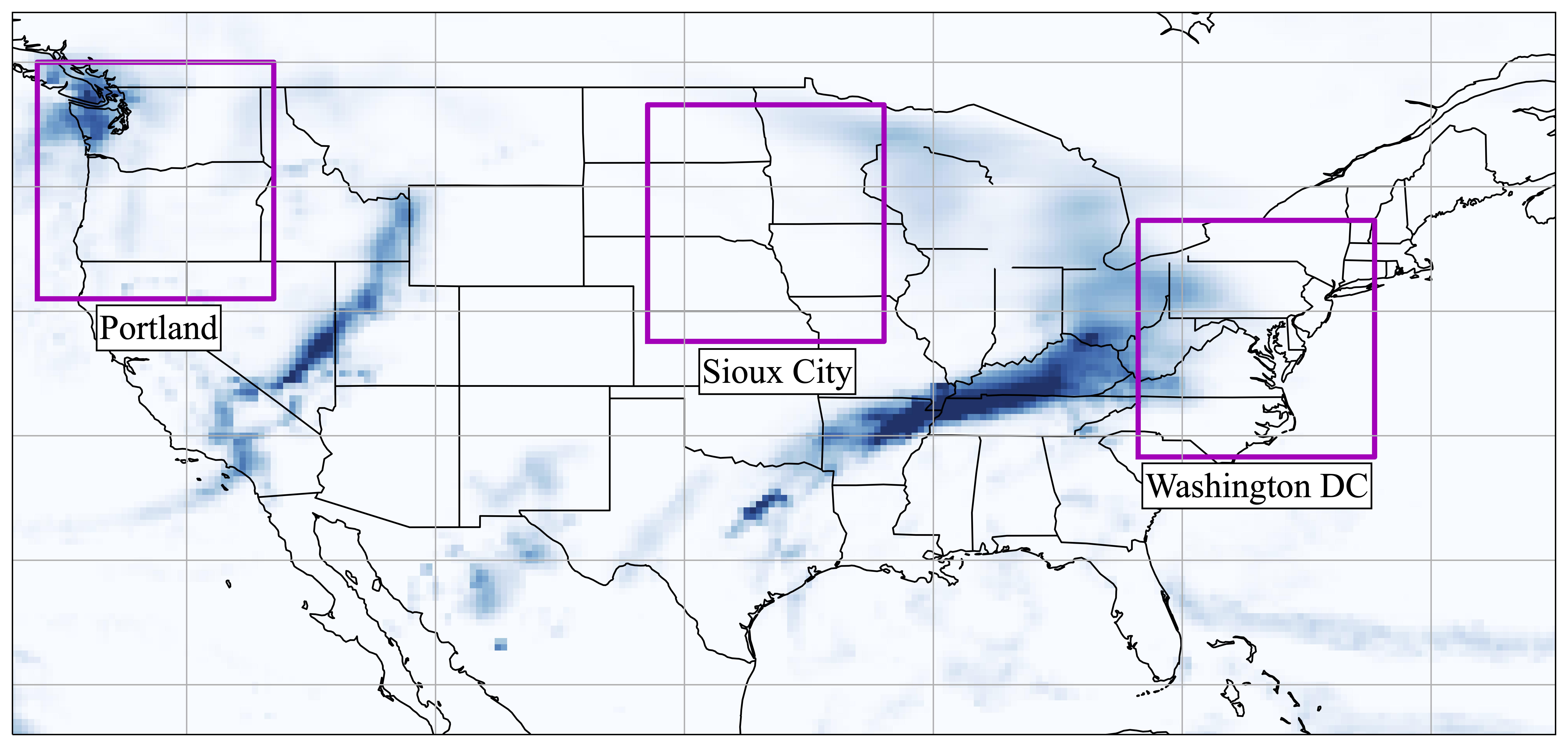}
\end{center}
\caption{The regions within which cGAN was applied, see also table \ref{regions:table}.
%
Model regions. $940 \times 940$ longitude-latitude grid points at $0.01^\text{o}$ resolution. Ranges start and end at the grid point centres.}
\label{cGAN_regions:fig}
\end{figure}

The model employed here is the conditional generative adversarial network ``cGAN'' developed in \citeA{Harris22} over the UK region. cGAN is in turn based upon a rainfall downscaling model developed by \citeA{Leinonen21}. In \citeA{Harris22}, the cGAN model takes ECMWF HRES \cite{IFS_CY47R3_partIII} forecasts of multiple atmospheric variables at a 0.1$^\text{o}$ resolution with lead times between 6 and 18 hours, and outputs an ensemble of rainfall predictions of NIMROD \cite{MetOffice03} rainfall radar data at the same time with a resolution of 0.01$^\text{o}$ ($\sim 1$km). In this study variables from the lower resolution ECMWF ERA5 reanalysis \cite{Hersbach20} and satellite determined orography are used as inputs (see table \ref{inputs:table}) and cGAN is trained to produce an ensemble forecast of the stage IV rainfall product \cite{Du11} linearly interpolated from $\sim 4$km to $\sim 1$km over three USA regions (figure \ref{cGAN_regions:fig}) and NIMROD over the UK.

\begin{table}
\centering
\begin{tabular}{p{50mm} p{75mm}}
\hline
%
\begin{hangparas}{.175in}{1}
1. Total Precipitation \newline
2. Convective Precipitation \newline
3. Total column liquid water \newline
4. Total column water vapour \newline
5. Surface Pressure \newline
6. Convective avaliable potential energy \newline
7. Zonal wind at 700 hPa \newline
8. Meridional wind at 700 hPa \newline
9. Total incoming solar radiation
\end{hangparas} & ECMWF ERA5 reanalysis variables linearly interpolated onto a 0.1$^\text{o}$ resolution longitude-latitude grid. $94 \times 94$ pixels. \\
\hline
10. Elevation & The terrain elevation at $0.01^\text{o}$ resolution. Derived from the GMTED2010 data set \cite{Danielson2010} for the regions in the USA and the same orography field used in \citeA{Harris22} for the UK, which is a approximately 1.25 km resolution field developed for a higher resolution version of the IFS. $940 \times 940$ pixels. \\
\hline
11. Land-sea mask & The fraction of permanent water bodies, including rivers, lakes and ocean. Derived from the ESA WorldCover 2020 dataset \cite{Zanaga2021} for the USA and derived from the IFS operational land-sea-mask as in \citeA{Harris22} for the UK. $940 \times 940$ pixels. \\
\hline
\end{tabular}
\caption{Model inputs}
\label{inputs:table}
\end{table}

\begin{table}
\begin{tabular}{|l|l|l|l|}
\hline
\bf{Region} & \bf{Centered on} & \bf{Latitude} & \bf{Longitude} \\
\hline
Atlantic Coast & Washington DC & 34.2$^\text{o}$N - 43.6$^\text{o}$N & 81.72$^\text{o}$W - 72.32$^\text{o}$W \\
Great Plains & Sioux City & 38.84$^\text{o}$N - 48.24$^\text{o}$N & 101.43$^\text{o}$W - 92.03$^\text{o}$W \\
Pacific North-West & Portland & 40.55$^\text{o}$N - 49.95$^\text{o}$N & 125.95$^\text{o}$W - 116.55$^\text{o}$W \\
UK & UK national grid & 49.55$^\text{o}$N - 58.95$^\text{o}$N & 7.45$^\text{o}$W - 1.95$^\text{o}$E \\
\hline
\end{tabular}
\label{regions:table}
\end{table}

Training was performed on a single NVIDIA A100 accelerator with reduced numerical precision (\ref{precision:sect}) taking around 4-5 days to train each cGAN model. As in \cite{Harris22} the data across split into smaller sub-images of 20 × 20 (low-resolution) and 200 × 200 (high-resolution), by randomly sampling patches from the full-sized images. In total, for each model, 640'000 samples were taken. For all regions except for Portland, the training data was taken from 2016, 2017 and 2018. For Portland the training data was taken from 2018 and 2019. As reported in \citeA{Harris22} the quality of the trained model does not converge smoothly. So evaluations of the CRPS against 2019 validation data (2020 for Portland) were performed on 33 models selected from the final third of the training run. The model with the lowest CRPS was then chosen as the model for that region. Results reported here were computed by evaluating this model on unseen 2020 test data for all regions except for Portland and 2021 test data for Portland.

An additional model was trained using data from all four regions resulting in almost four times the quantity of training data. The model trained on all regions used all of the training, validation and test data, still segregated into training/validation/test sets, as described above.

\section{Data}
\subsection{Nimrod}

Over the UK we use the NIMROD 1-km data product \cite{MetOffice03}. A number of radar stations across the UK and Ireland provide data at 5 minute intervals, which is then processed to calibrate, correct for hail and remove artefacts. Close to the radar stations the spatial resolution is around 1 km and reduces to around 5km further away. This data is then merged onto a 1 km resolution national grid which includes regions where the true resolution is worse than 1km. To process the NIMROD data for cGAN, a sub-region of 5 minute snapshots of rainfall rates are averaged over 1 hour periods and linearly interpolated to a 0.01$^\text{o}$ longitude-latitude grid. The procedure is identical to that employed by \citeA{Harris22}.

\subsection{Stage IV}

NCEP/EMC Stage IV Data is a gridded rainfall data set over the USA at $\sim4$ km resolution and hourly time intervals. It is derived using a combination of rain gauges, rainfall radar by the 12 River Forecast Centers in the continental USA who use differing algorithms and apply local manual quality control. It is then mosaicked together by NCEP, \cite{Nelson16}. An eastern portion of the stage IV data is operationally assimilated into IFS \cite{Lopez11}

\subsection{Terrain and Land-sea mask}

Elevation data in the USA is derived from the 30 arc-second, ($\sim 1$ km at the equator) GMTED2010 data set \cite{Danielson2010}. In each of the 3 USA regions is it interpolated, using the nearest neighbour, onto a $0.01^\text{o}$ longitude-latitude grid. The land sea mask is derived from the 10m resolution ESA WorldCover 2020 dataset \cite{Zanaga2021}. Each of the elevation grid points corresponds to a subset of nearest neighbour WorldCover grid points. The land sea mask is the fraction of these points that are of permanent water bodies, including rivers, lakes and ocean. Elevation data for the UK is unchanged with respect to the model used by \citeA{Harris22}. It consists of a $\sim 1.25$ km resolution elevation dataset developed for very high resolution versions of IFS and the land-sea mask used for the operational HRES forecast.

\subsection{ERA5}

For practical convenience, in this study we have substituted the short range IFS forecast used for training cGAN in \citeA{Harris22} for the ERA5 reanalysis \cite{Hersbach20}. Both systems use the same family of data assimilation algorithms to obtain their fields, however ERA5 uses an older version with a lower resolution than the operational system. ERA5 might however benefit from more data that didn't make it into the operational model in time. In the operational IFS forecast it has been suggested that initial rainfall predictions take some time to ``balance'', which is part of the motivation for using forecasts out to 1 day instead of the initial condition. We don't know if this is also a problem with ERA5.

We would therefore expect that training using ERA5 would lead to broadly the same conclusions as if we trained with the IFS forecast, however in an operational setting, it might me more optimal to use the best model available.

\section{Scores}

To compare against the IFS ensemble forecast we focus on three metrics used in \citeA{Harris22}, namely the Root-Mean-Squared Error of the Ensemble Mean (RMSEEM), the Continuous Ranked Probability Score (CRPS) \cite{Gneiting07}, the Radially Averaged Log Spectral Distance (RALSD) and add a fourth, the variogram score \cite{Scheuerer15}.

The Root-Mean-Squared Error of the Ensemble Mean (RMSEEM) has the advantage that it is simple and quantifies the ability of an ensemble forecast to represent the mean of the distribution. In contrast, the Root-Mean-Squared error of an individual forecast is particularly problematic for rainfall. This is because rainfall events can be quite local and intense. If a rainfall event is correctly forecast, but it is in slightly the wrong place the RMS error of an individual forecast might be higher than not forecasting rainfall at all. It is therefore not included.

The Continuous Ranked Probability Score (CRPS) \cite{Gneiting07} quantifies the quality of the entire forecast distribution independently at each grid point, but completely ignores the relation between grid points. It is a {\it proper score}, meaning that it is minimised when the forecast has the same distribution as the measurements.

Both the RMSEEM and CRPS can be used to assess the quality of the forecast at each point separately. Neither address the spatial structure of a forecast. The energy score \cite{Gneiting07} is a multi-dimensional generalisation of the CRPS \cite{Hersbach00}. However it lacks statistical power to distinguish covariances \cite{Pinson13} and therefore the spatial structure of forecasts. The fractions skill score \cite{Roberts08,Roberts08b} is a popular score for spatial verification. However it can be difficult to interpret (\ref{FSS:sect}) specifically in regions or times of low rainfall  \cite{Mittermaier21}.  Instead we focus on the RALSD because it was used in \citeA{Harris22} and the variogram score because it was developed specifically to address these problems.

\subsection{The Radially Averaged Log Spectral Distance (RALSD)}

The Radially Averaged Log Spectral Distance (RALSD) is a score used by \citeA{Harris22} to measure the quality of the spatial relationship between forecast locations:
\begin{equation}
\text{RALSD} = \sqrt{ \frac{1}{N} \sum_{i=1}^N \left( 10 \log_{10} \overline{P}_{\text{radar},i} - 10 \log_{10} \overline{P}_{\text{model},i} \right)^2 }
\label{RALSD:eqn}
\end{equation}
Here $\overline{P}_{\text{model},i}$ and $\overline{P}_{\text{radar},i}$ are the radially averaged power spectra of the respective model and radar data and $N$ is the number of points in the spectra after radial averaging. In \citeA{Harris22} $\overline{P}_{\text{model},i}$ and $\overline{P}_{\text{radar},i}$ were computed for single images and then the RALSD for each image was averaged over all ensemble members and sample dates. Here we find $\overline{P}_{\text{model},i}$ and $\overline{P}_{\text{radar},i}$ by averaging radially and over our data set including ensemble members and different rainfall dates, before computing the RALSD. In our case some regions are irregular, most notably the Portland region, see eg. figure \ref{CRPS:fig}. In order to compute $\overline{P}_{\text{radar},i}$ values are set to zero in the masked region within the bounding rectangle. This will introduce some inaccuracy in this score and so the same inaccuracy is introduced into the forecasts for the calculation of $\overline{P}_{\text{model},i}$, by also setting the equivalent regions to zero. The power spectrum is the discrete Fourier transform of the discrete covariance function which leads us to the variogram score which can be computed without this inaccuracy.

\subsection{The variogram score}

The variogram score \cite{Scheuerer15} is designed to assess the forecast representation of the relation between variables. It is a proper score because is it minimised when the forecast distribution equals the true distribution. It is not strictly proper because there are other distributions that can match the score of the true distribution. The variogram score measures the difference between variograms which are closely related to the covariance and correlation. The variogram may be defined as
\begin{align}
2 \gamma_{ij} \left( X_i, X_j \right)
&= \text{Var} \left[ X_i - X_j \right] \\
&= \text{E} \left[ \left( X_1-X_2 \right)^2 \right] - \text{E} \left[ X_i-X_j \right]^2 \\
&= \text{Var} \left[ X_i \right] + \text{Var} \left[ X_j \right] - 2 \text{Cov} \left[ X_i,X_j \right].
\end{align}
Here $X_i$ and $X_j$ represent a random process at two locations $i$ and $j$. The variogram score for a forecast is defined as
\begin{equation}
S(\mathbf{y},\mathbf{X}) =
\sum_{i=1}^{d-1}
\sum_{j=i+1}^d w_{ij}
\left( \left|y_i-y_j\right|^p - E_F \left[ \left| X_i-X_j \right|^p \right] \right)^2
\label{variogramScore:eqn}
\end{equation}
where $d$ is the number of pixels in each image, $y_i$ denotes a measurement at pixel $j$ and $E_F \left[ \left| X_i-X_j \right|^p \right]$ denotes the expectation of the difference between the forecast $\mathbf{X}$ at pixels $i$ and $j$. When the forecast distribution is in the form of an $m$ member ensemble $\mathbf{x}^{(1)},\mathbf{x}^{(2)},\dots,\mathbf{x}^{(m)}$
\[
E_F \left[ \left| X_i-X_j \right|^p \right] \approx \frac{1}{m} \sum_{k=1}^m \left| x_i^{(k)}-x_j^{(k)} \right|^p, \qquad i,j=1,2,\dots,d.
\]
$p$ is a parameter which we choose $p=0.5$, which appears to be ``good'' for multivariate normal distributions \cite{Scheuerer15}. $w_{ij}$ is a selection of user defined weights that correspond to how important each pair of points is. In the estimate of the expectation values, the difference between two points often increases with distance because their correlation decreases. Down-weighting pairs that are expected to have relatively weak correlations can therefore benefit the signal-to-noise ratio. We therefore choose
\[
w_{ij} = \left\{ \begin{array}{ll}
\exp(-k D(i,j)) & D(i,j) \leq D_\text{max} \\
0 & \text{otherwise}.
\end{array} \right.
\]
where $D\left( i,j \right)$ is function that returns the distance in pixels between points $i$ and $j$ so that more distant relationships are considered to be less important. $D_\text{max}$ is a cut-off distance in pixels and $k$ is a decay constant that sets the decay in $w_{ij}$ to approximate the decay in rainfall spatial auto-correlation with distance in pixels.

The double sum in equation (\ref{variogramScore:eqn}) makes computation of the variogram score expensive over large images. When the region is regular it is possible to speed this up using the Fast-Fourier-Transform. However, in our case the regions are irregular. We therefore compute the variogram score over low resolution ($94 \times 94$ pixels $\sim 10$ km resolution) spatial averages of the rainfall. This choice can be justified by the fact the spatial auto-correlation changes relatively slowly over high resolution ($\sim 1$km) pixels. We choose $D_\text{max}=5$ low resolution pixels ($\sim 50$km) corresponding to when the rainfall spatial auto-correlation falls to $\sim 0.4$, and $k = 0.175$ $(\text{low resolution pixels})^{-1}$.

\section{Results}

All results are computed from 256 randomly chosen dates and times from the test data year. We use 50 ensemble members, in contrast to the 100 used in \citeA{Harris22}, because we compare to the IFS ensemble, that also has 50 members. Scores such as the CRPS have biases that are a function of ensemble size. To fairly compare the CRPS of cGAN and interpolated IFS we therefore need to use the same ensemble size.

\begin{figure}
\begin{center}
\includegraphics[width=\textwidth]{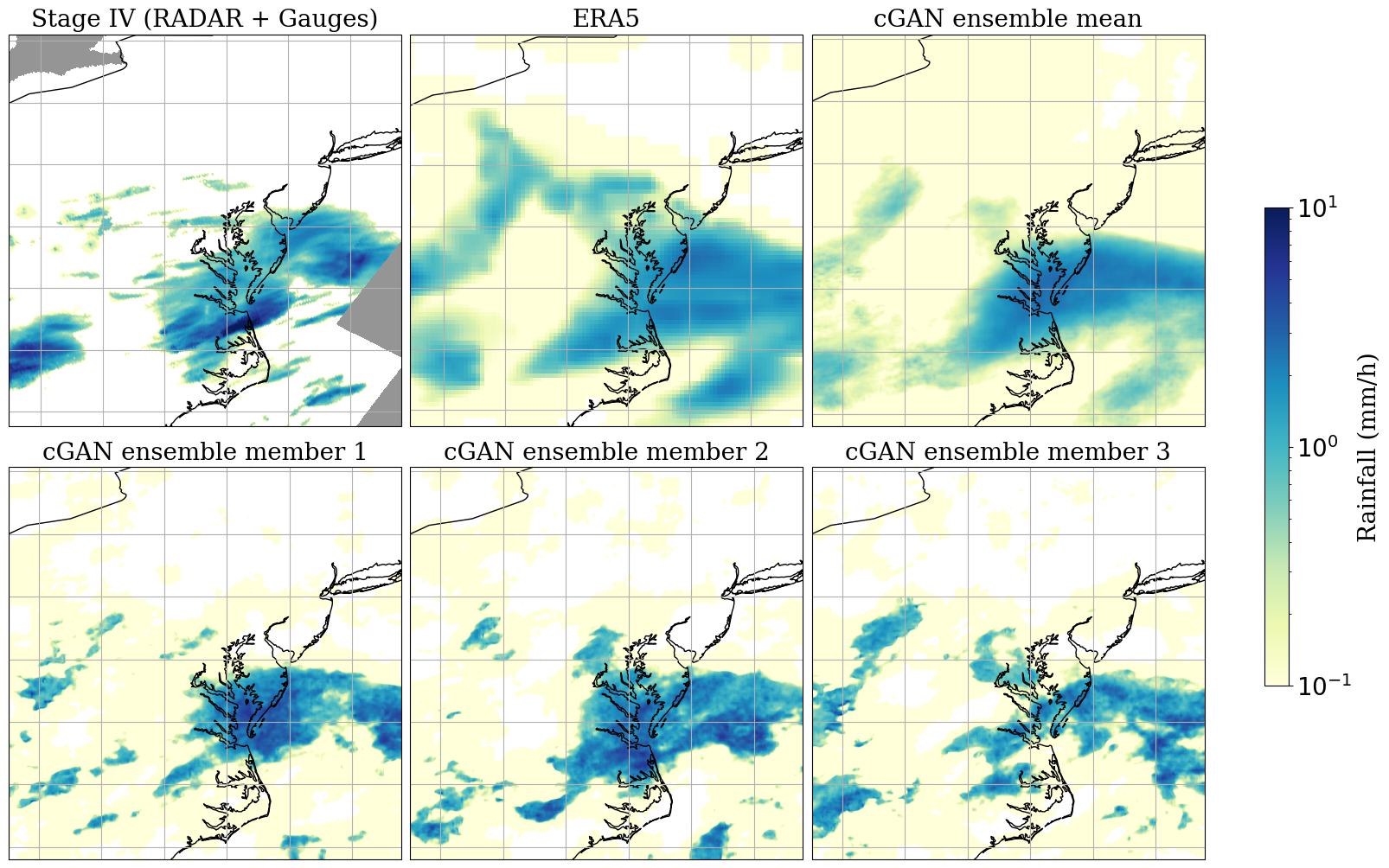}
\end{center}
\caption{Three example cGAN ensemble members in the Atlantic coast region entered on Washington DC on 14th January 2020 at 15:00 UTC. The grey regions in the radar plot correspond to no stage IV data. Although cGAN makes predictions in the grey regions, these are not included when evaluating scores.}
\label{Example:fig}
\end{figure}

Figure \ref{Example:fig} provides an example of cGAN inputs and outputs for a particular time: The Stage IV data (radar + gauges) is what we use as the ``truth'' and is not seen by cGAN. The inputs to cGAN come from ERA5 and other variables, see table \ref{inputs:table}. In figure \ref{Example:fig} it is clear that ERA5 rainfall is in a different location to Stage IV and looks blurry due to it's lower resolution. The cGAN ensemble mean prediction has a similar intensity to the ERA5 rainfall and is in a different location, that although approximately covers the region of Stage IV rainfall, is not in exactly the same place. Each ensemble member prediction, along the bottom row, displays an intensity of rainfall much more similar to Stage IV, although very low rainfall still seems to be over predicted. Note that the somewhat horizontal bands of rainfall, present in Stage IV, but not present in the ERA5 rainfall, are represented in the cGAN predictions, showing the ability of cGAN to predict the spatial structure of rainfall patterns. 

\subsection{CRPS and RMSEEM scores}
\label{CRPS:sect}

The scores in table \ref{CRPS:table} indicate that with respect to the CRPS, each model performs well relative to the IFS ensemble, in its own region. That is, from ERA5 data cGAN creates a rainfall ensemble that is competitive with the IFS ensemble at short lead times. The model that was trained on all regions performed even better.

Examining how well models trained in one region predict another indicates how well generic systematic biases and uncertainty in the ERA5 rainfall model are corrected in cGAN. The Sioux City model for example, which is the model trained using data exclusively from the Sioux region, had a low CRPS in the Sioux region. However, it appears to be particularly poor at forecasting elsewhere. The Washington model, in contrast, appears to give a good CRPS everywhere, including in the Sioux City region. We have a few hypothesis for why the Washington model is doing so well. Firstly, it is possible that the Washington region has all of the ingredients necessary to fit cGAN to generic weather. Much like the UK, there is variable elevation, including areas of ocean. The UK model also does relatively well on all regions, however not as well as IFS or the Washington model. There is also a lot of variability over both the Washington and UK regions, in contrast to the Sioux and Portland regions where there are long periods or areas of dry weather. So it might also be that the Washington model has effectively more data to train on. The scores of the Washington model motivated the production of the all region model which equaled or outperformed the other models.

The model ``cGAN UK HRES'' refers to the original model trained in \citeA{Harris22} which used IFS HRES forecasts as its input instead of ERA5. We used this model, without any re-training, with ERA5 data as its input instead, which might be expected to break it. However, it outperforms the ``cGAN UK'' model that is identical except that it was trained using ERA5. The number in brackets in table \ref{CRPS:table} is the score reported by \citeA{Harris22} using HRES inputs. This suggests that using a better dynamical forecast model results in improved cGAN predictions, in both training and evaluation.

The ``IFS ensemble'' model represents the output of the entire 50 member ensemble, linearly interpolated to the 1km grid, effectively the simplest downscaling method. The mean-absolute-error of IFS ensemble member 2 illustrates the improvements to the CRPS in this context by moving from a deterministic to an ensemble forecast. The higher resolution HRES deterministic forecast outperforms ensemble member 2, however it still does not do better than the full ensemble by this metric. Linearly interpolating ERA5, results in similar CRPS to the HRES forecast in our context. ERA5 being the input data, these numbers show how much improvement cGAN is able to make.

A known problem with rainfall forecasts is that they tend to drizzle with light rain rather more than in reality and under represent the extreme rainfall events. To address this in the simplest way, the ERA5 rainfall was scaled to reproduce the Stage IV and NIMROD rainfall distribution. A given ERA5 rainfall quantity has a corresponding NIMROD rainfall quantity at the same point in its cumulative distribution function (CDF). This mapping was performed for each 1km pixel separately because although the CDF is less certain than for using all locations at once, it varies considerably in space, for example due to altitude. The resulting CRPS is improved for all regions, except for the UK, but not enough to fully explain, or outperform cGAN.

Predicting zeros everywhere actually does quite well on the CRPS, outperforming many of the deterministic forecasts. We are not sure why this is, since the issues of correctly predicting rainfall in randomly slightly the wrong place are not necessarily present in the CRPS. It might be due to the extreme nature of rainfall not being represented fully in the forecasts, or spatial biases in the rainfall distribution. The IFS ensemble and all cGAN models outperform predicting zero rainfall.

\begin{table}
\begin{center}
\begin{tabular}{l|c|cccc}
\toprule
& & & Data & &  \\
Model        & Metric & Portland &   Sioux  & Washington &    UK \\
\midrule
cGAN Portland      & CRPS & 0.068 & 0.106 & 0.278  & 0.231 \\
cGAN Sioux           & CRPS & 0.069 & 0.060 & 0.137  & 0.120 \\
cGAN Washington & CRPS & 0.064 & {\bf 0.057} & 0.113 & 0.101 \\
cGAN UK               & CRPS & 0.083 & 0.080 &  0.139  & 0.097 \\
cGAN UK HRES   & CRPS & 0.073 & 0.060 & 0.122 & {\bf 0.096} (0.086) \\
cGAN all regions.   & CRPS & {\bf 0.060} & {\bf 0.058} & {\bf 0.109} & 0.098 \\
IFS Ensemble.       & CRPS & 0.070 & 0.060 & 0.123 & 0.098 \\
IFS Ens. member 2 & MAE  & 0.102 & 0.095 & 0.188 & 0.144 \\
IFS HRES              & MAE  & 0.101 & 0.088 & 0.185 & 0.141 \\
ERA5                       & MAE & 0.096 & 0.086 & 0.181 & 0.155 \\
ERA5 PDF mapped & MAE & 0.086 & 0.088 & 0.175 & 0.162 \\
Zeros                       & MAE & 0.097 & 0.071 & 0.147 & 0.131 \\ 
\bottomrule
\end{tabular}
\caption{CRPS (lower is better) of cGAN trained on and generating an ensemble of rainfall forecasts using ERA5 data in all cases except cGAN UK HRES that was trained on and uses HRES data. In a deterministic forecast, each ensemble member is identical and the resulting CRPS is equal to the mean-absolute-error (MAE) which is also included. A description of the deterministic MAE models is in section \ref{CRPS:sect}.}
\label{CRPS:table}
\end{center}
\end{table}

\begin{figure}
\begin{center}
\setlength\extrarowheight{-20pt}
\addtolength{\tabcolsep}{-7pt}
\begin{tabular}{p{0.3\textwidth} p{0.3\textwidth} p{0.4\textwidth}}
  \vspace{0pt} \includegraphics[width=0.29\textwidth]{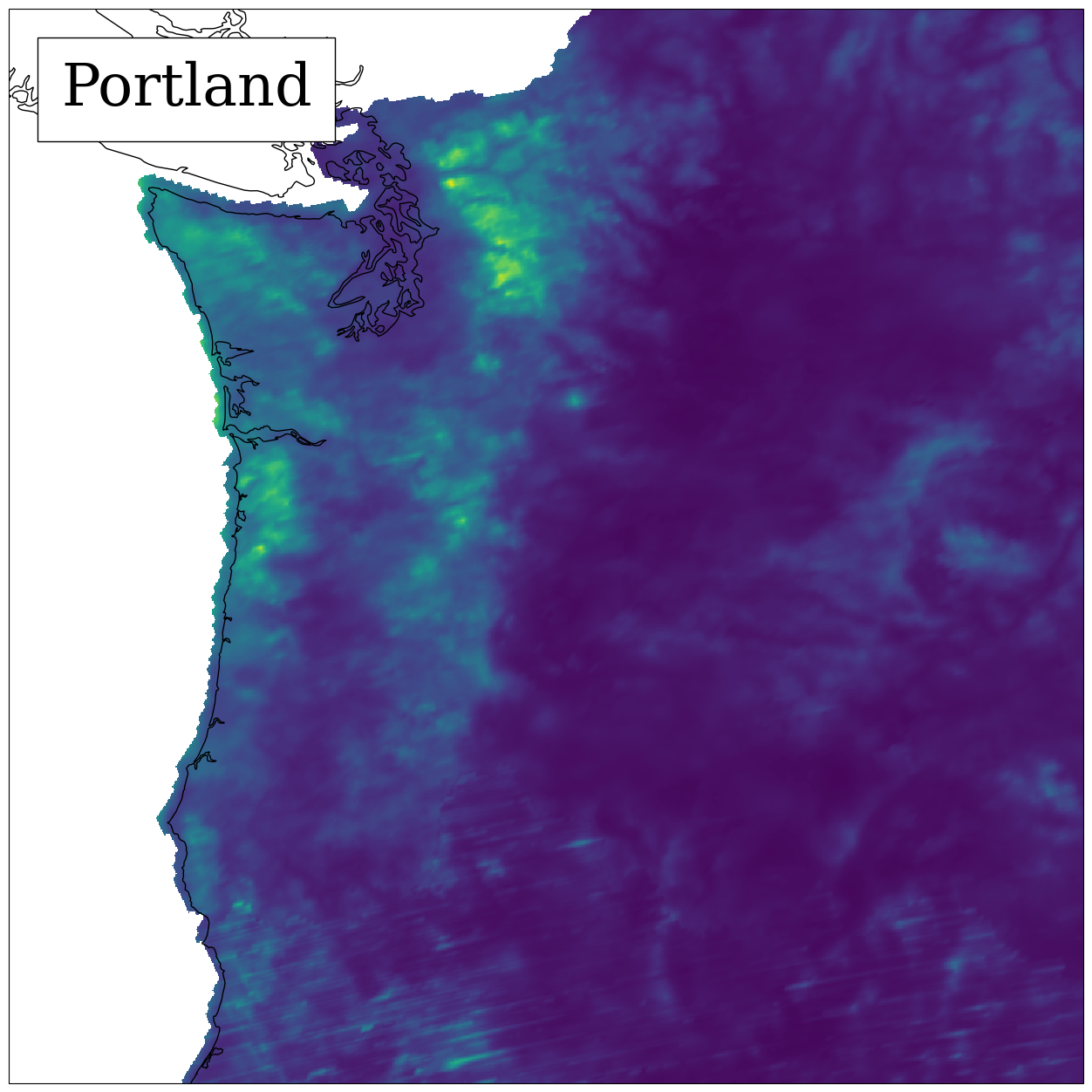} &
  \vspace{0pt} \includegraphics[width=0.29\textwidth]{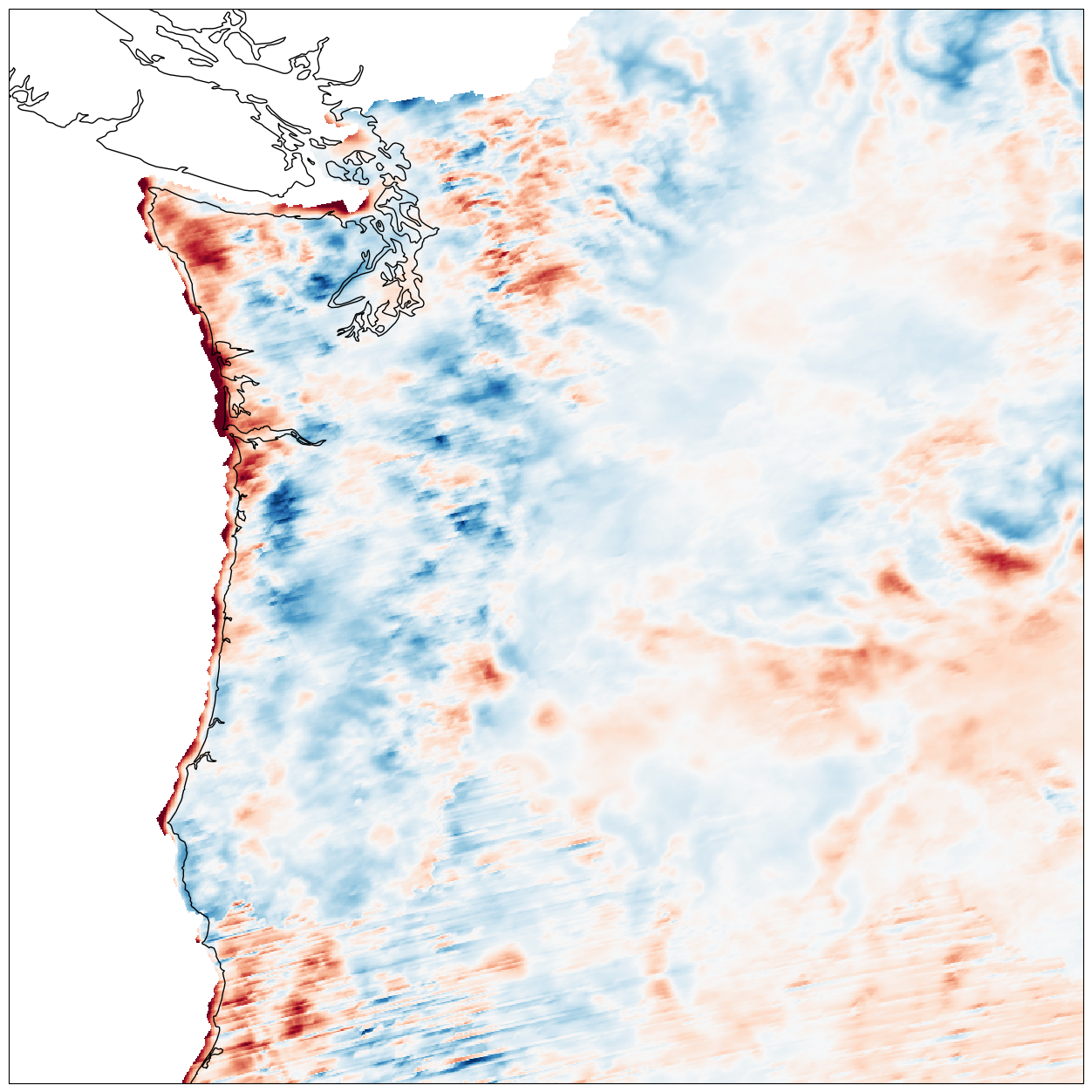} &
  \vspace{0pt} \includegraphics[width=0.29\textwidth]{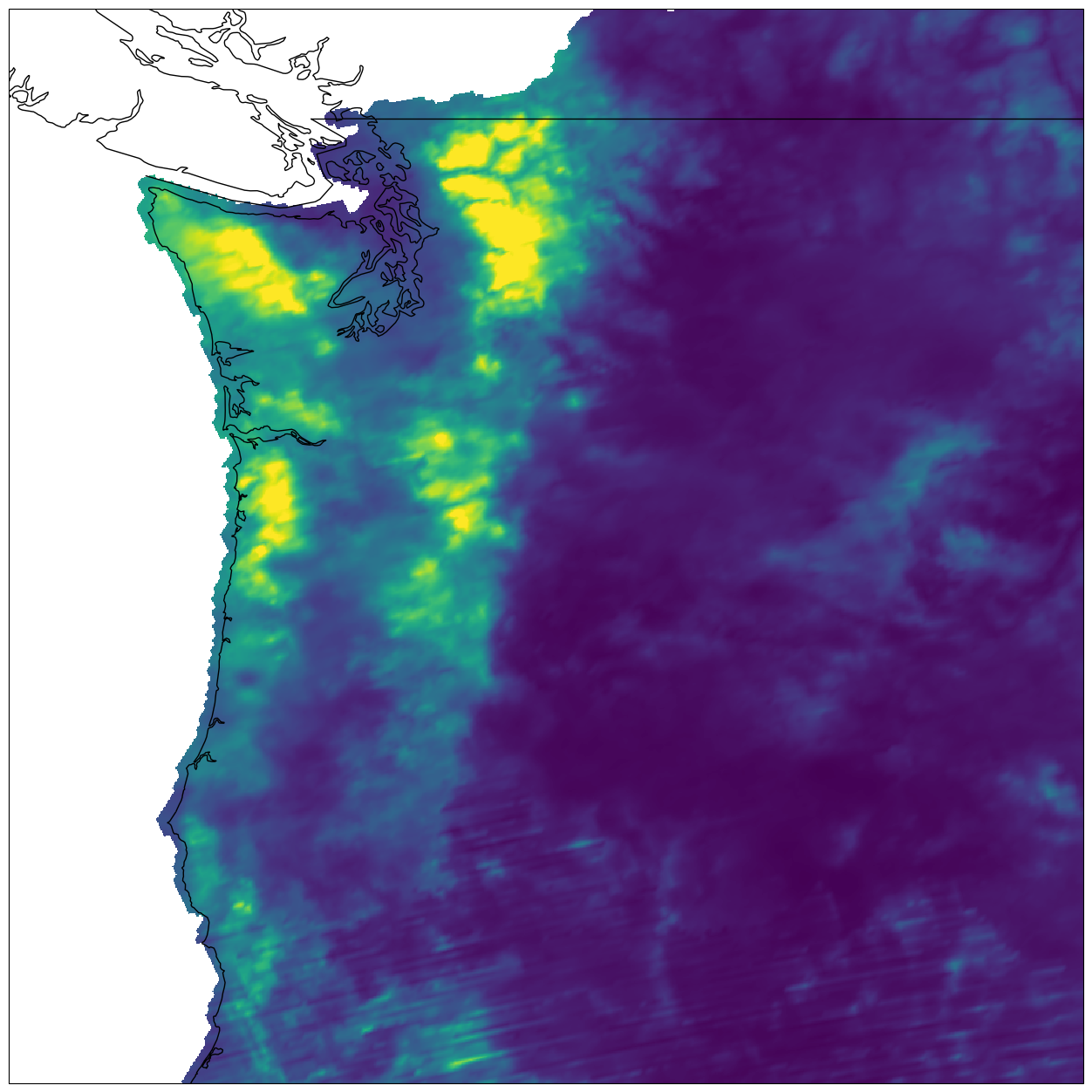} \\
  \vspace{0pt} \includegraphics[width=0.29\textwidth]{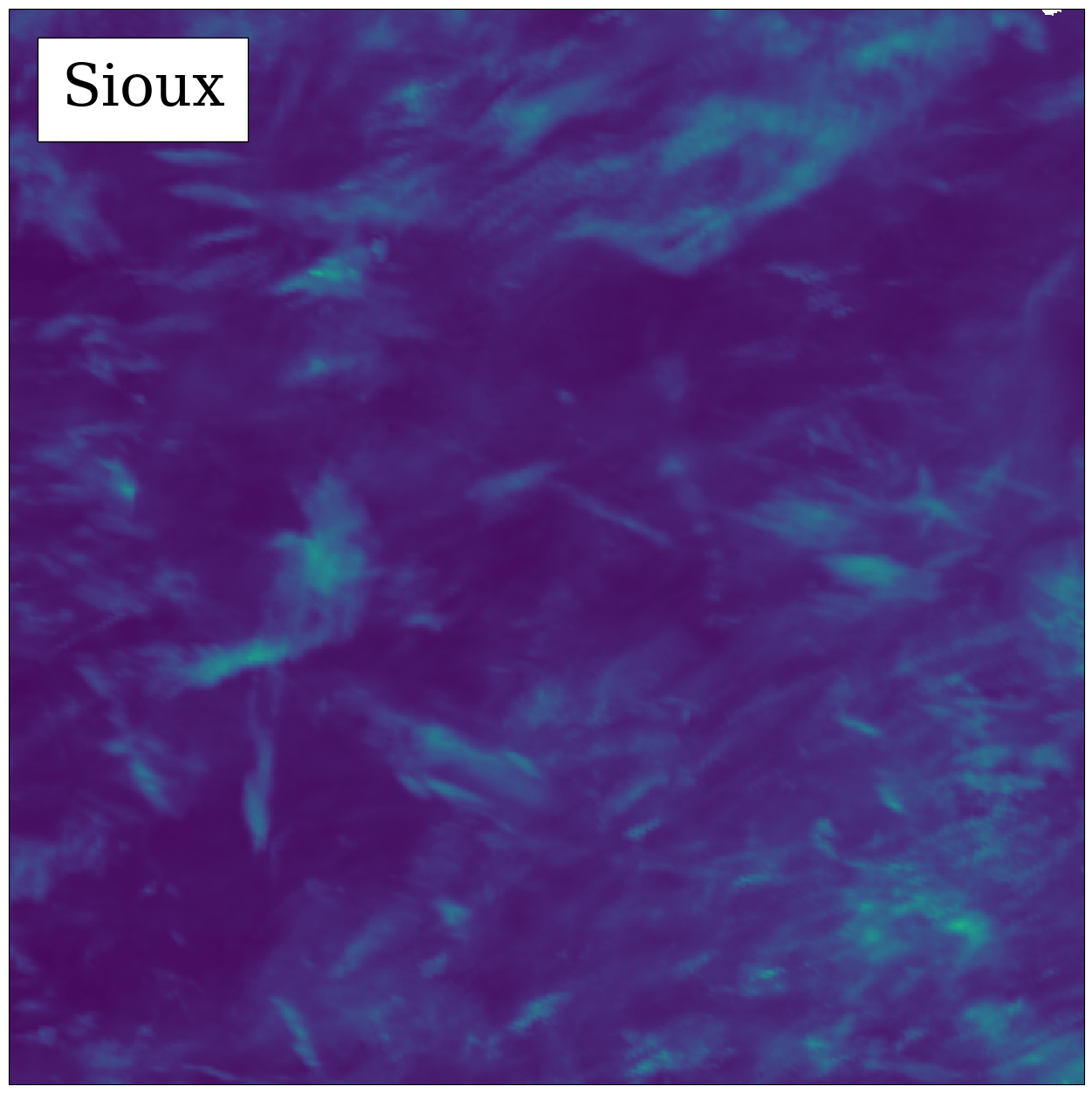} &
  \vspace{0pt} \includegraphics[width=0.29\textwidth]{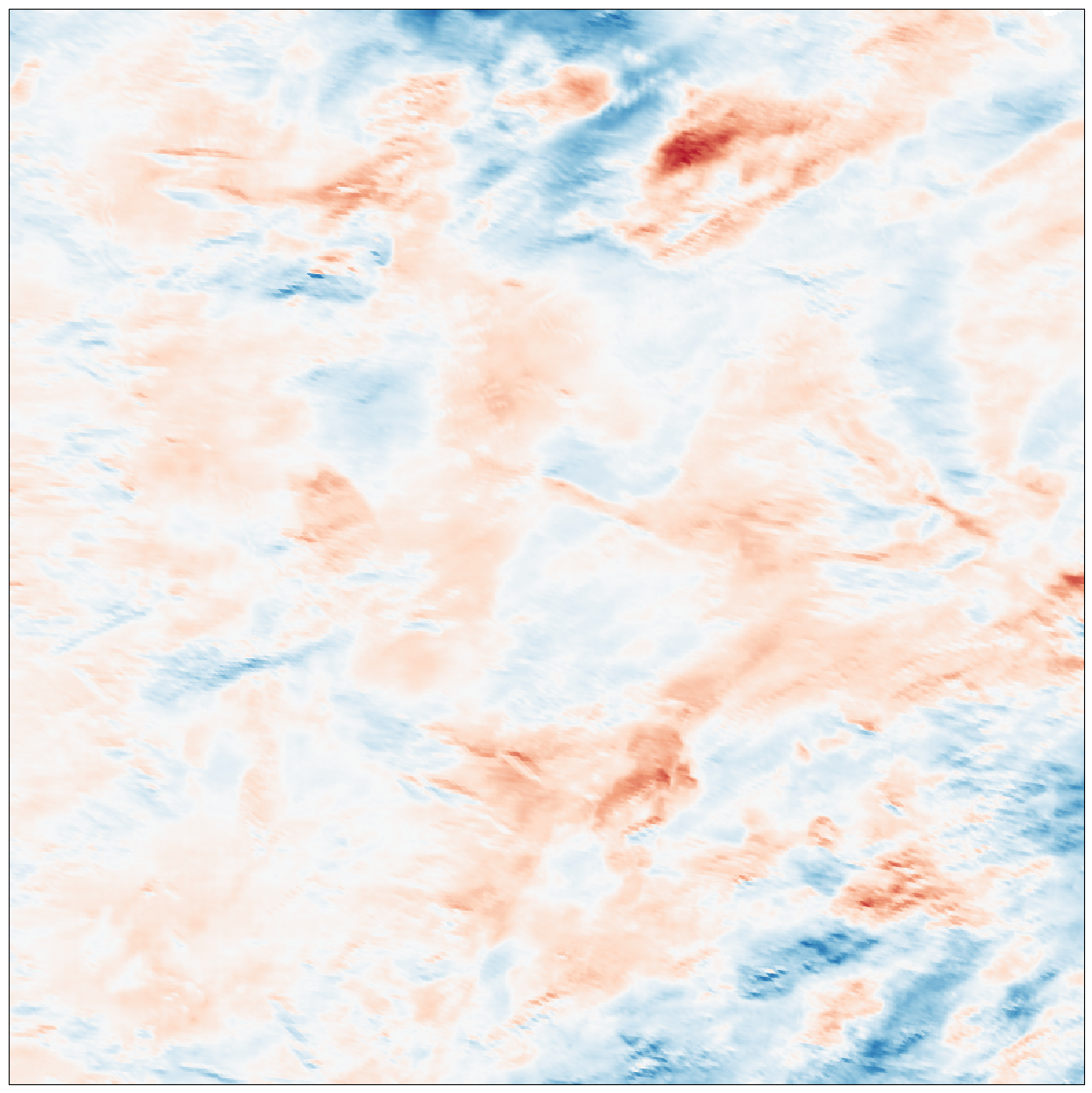} &
  \vspace{0pt} \includegraphics[width=0.29\textwidth]{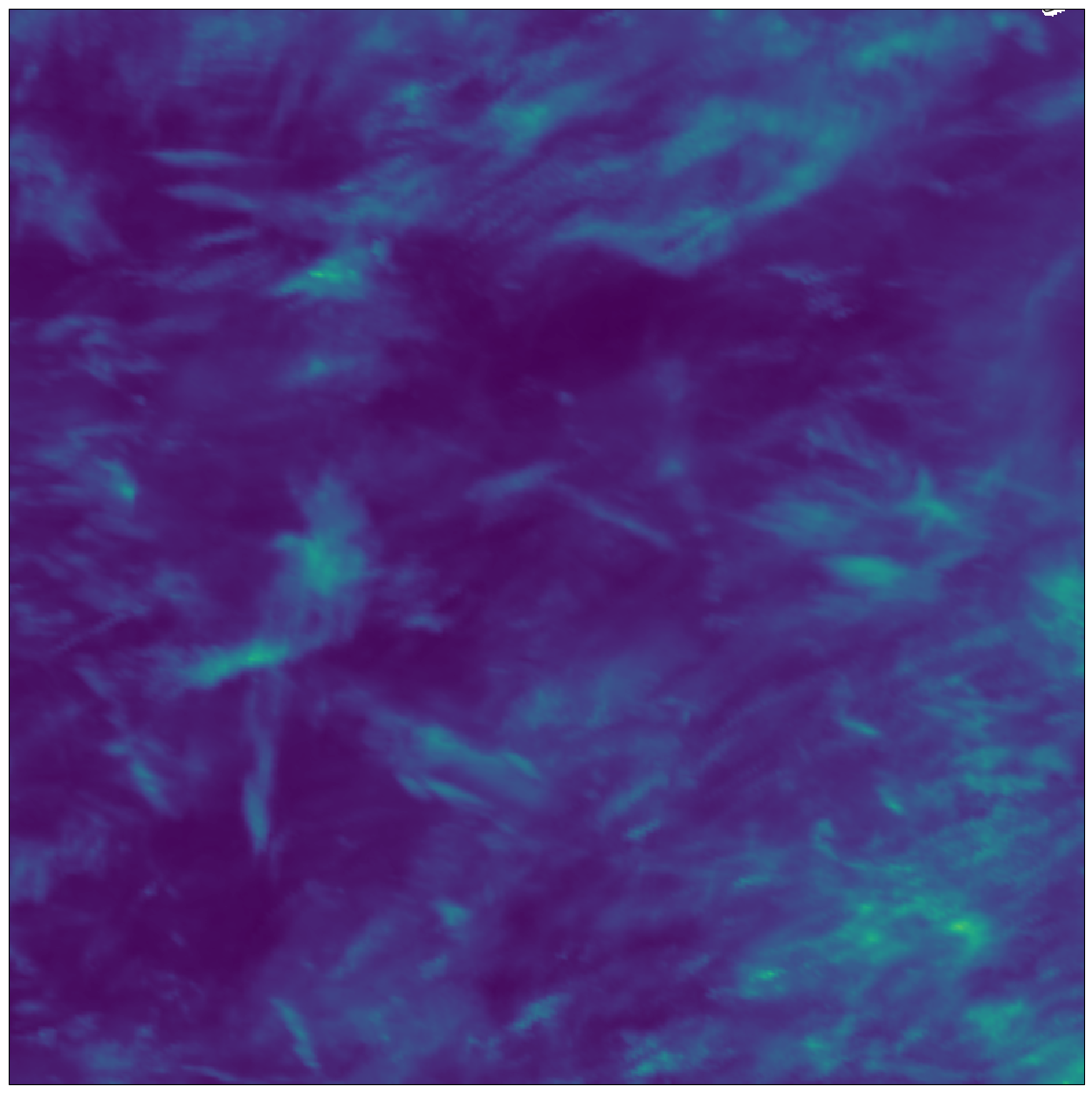} \\
  \vspace{0pt} \includegraphics[width=0.29\textwidth]{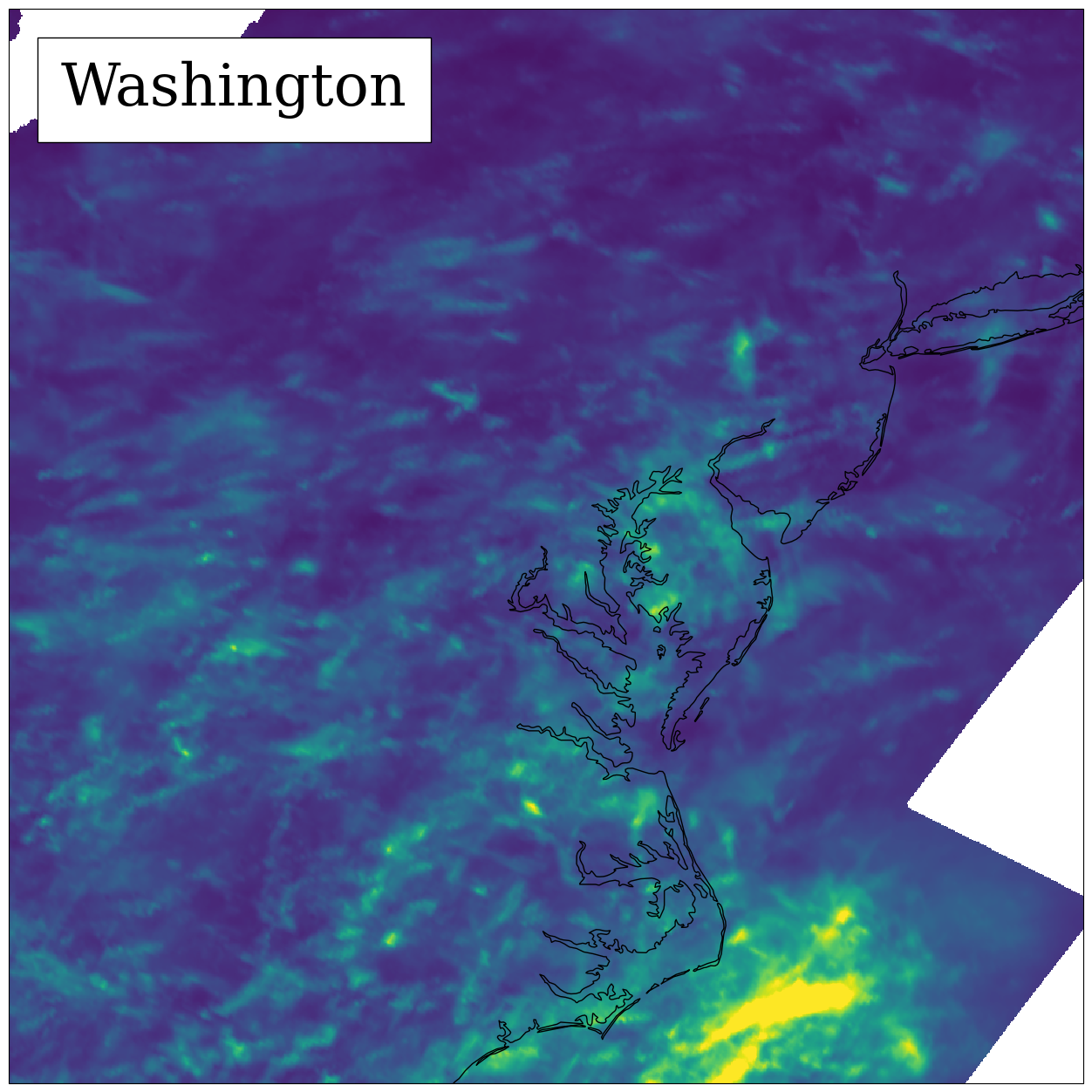} &
  \vspace{0pt} \includegraphics[width=0.29\textwidth]{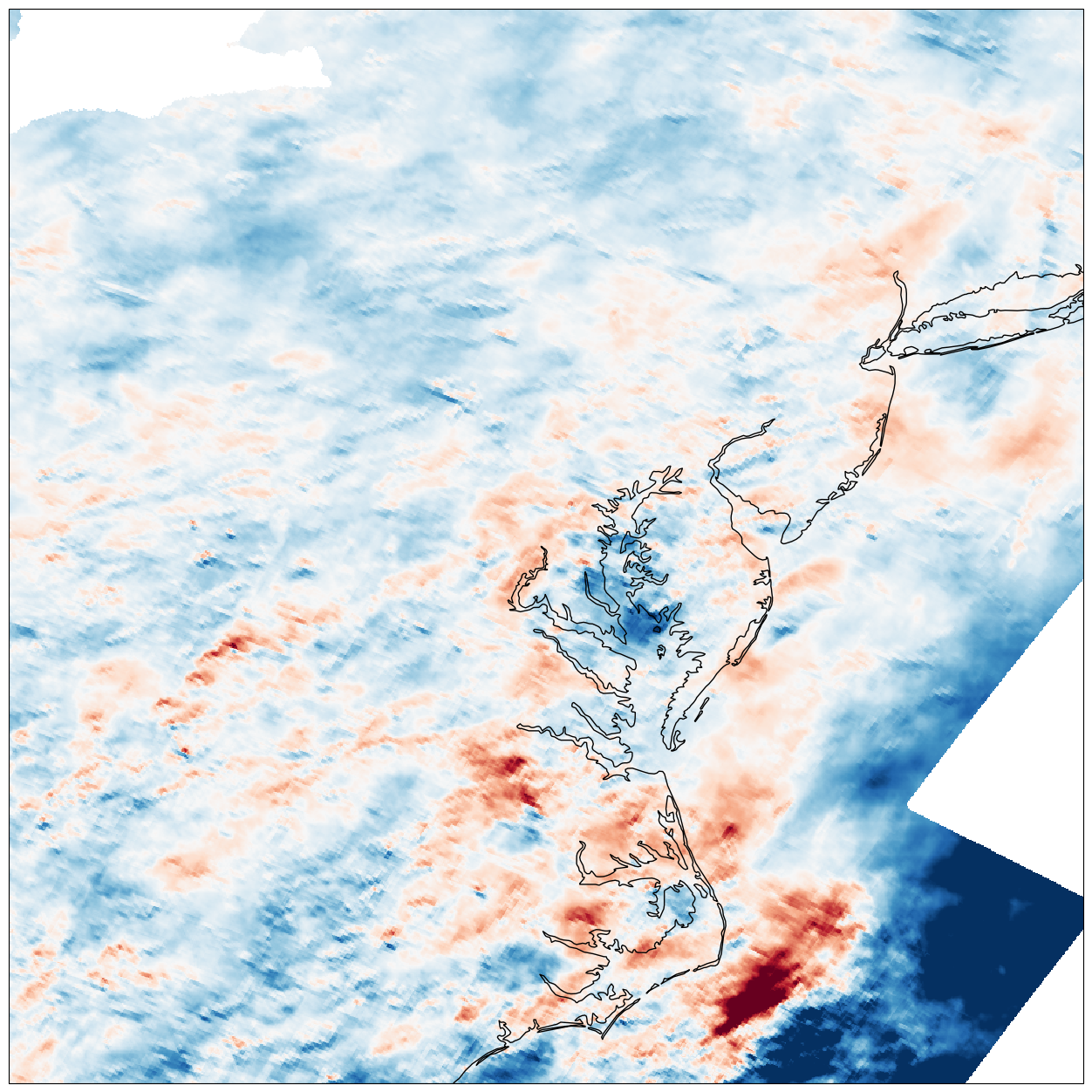} &
  \vspace{0pt} \includegraphics[width=0.29\textwidth]{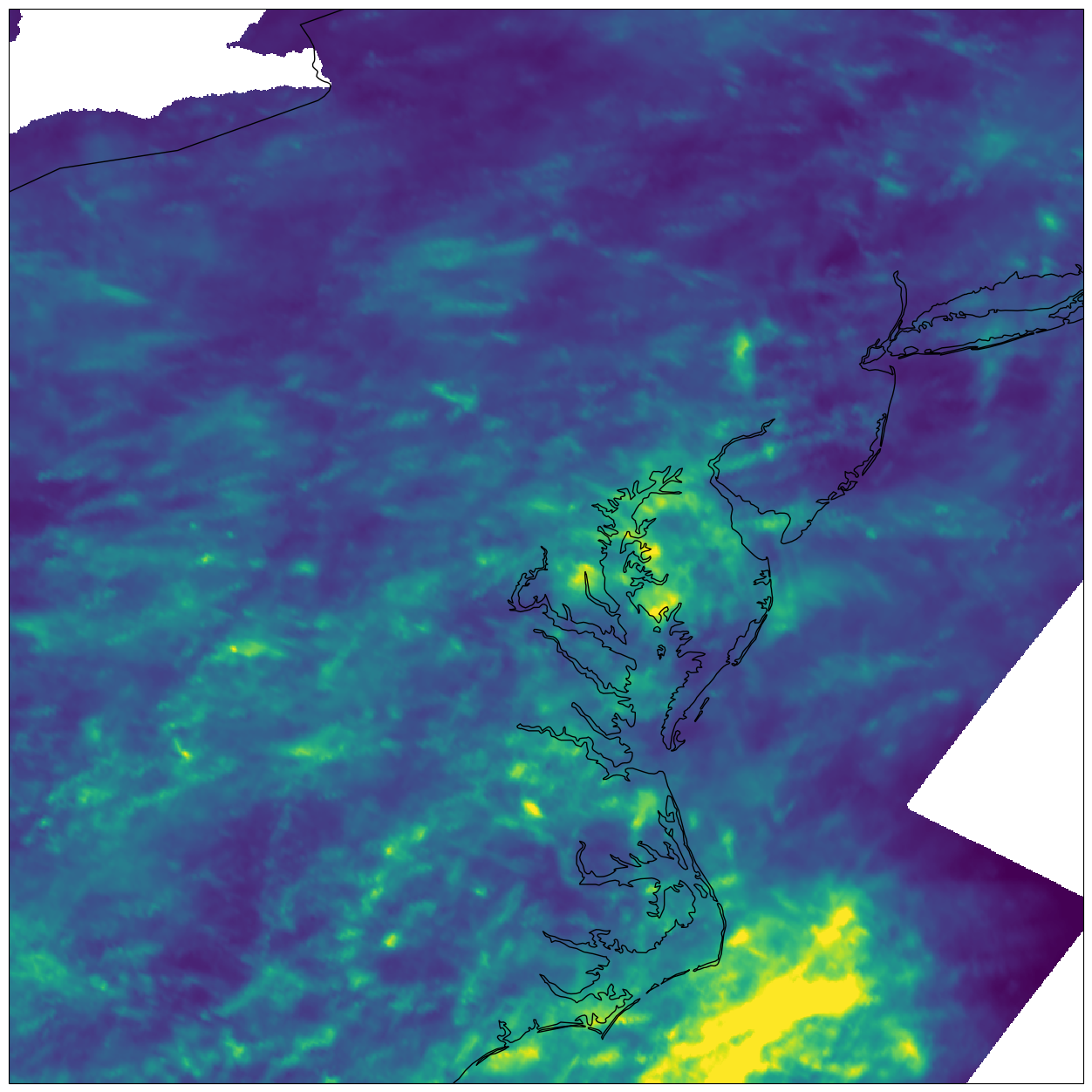} \\
  \vspace{0pt} \includegraphics[width=0.29\textwidth]{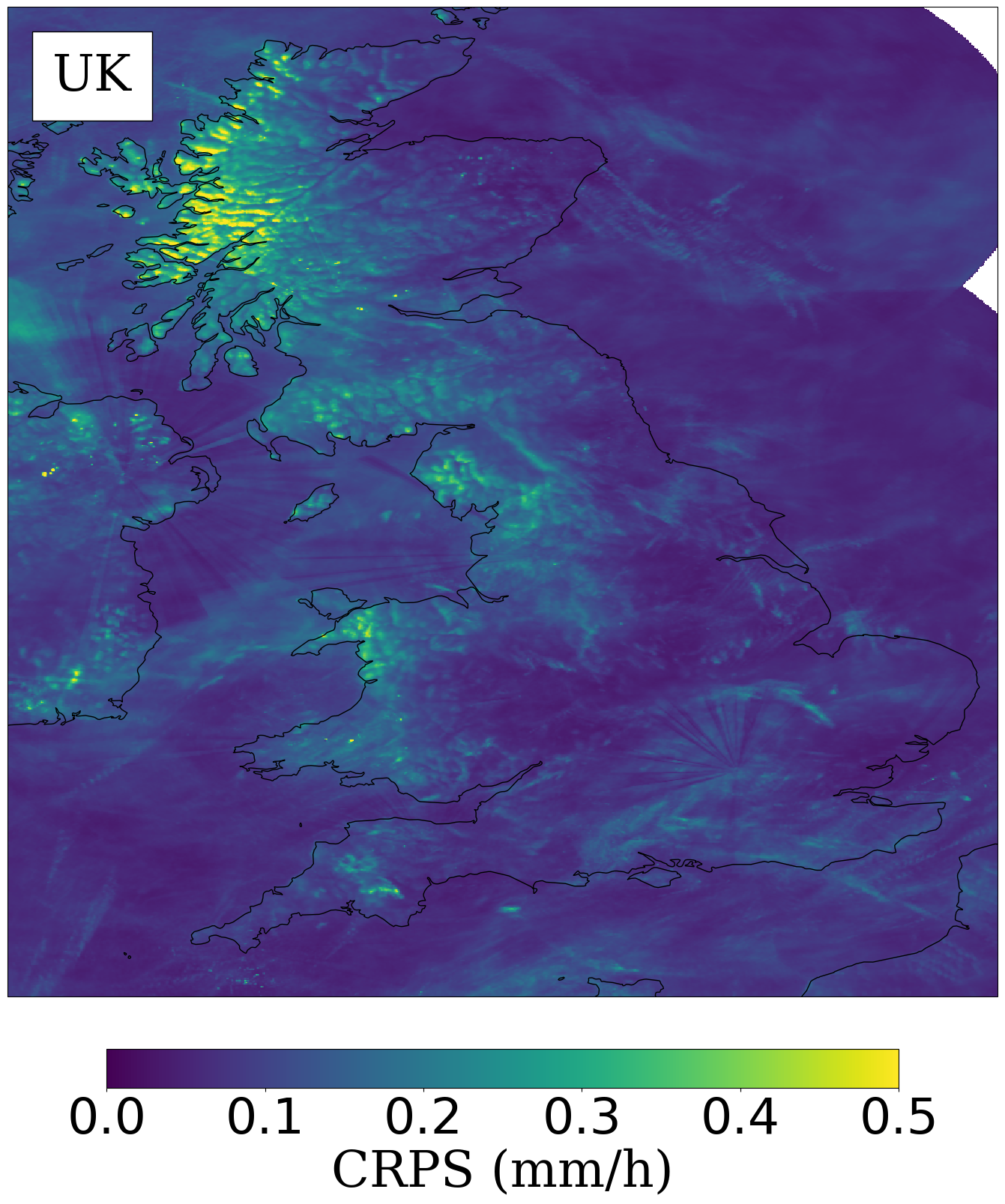} &
  \vspace{0pt} \includegraphics[width=0.29\textwidth]{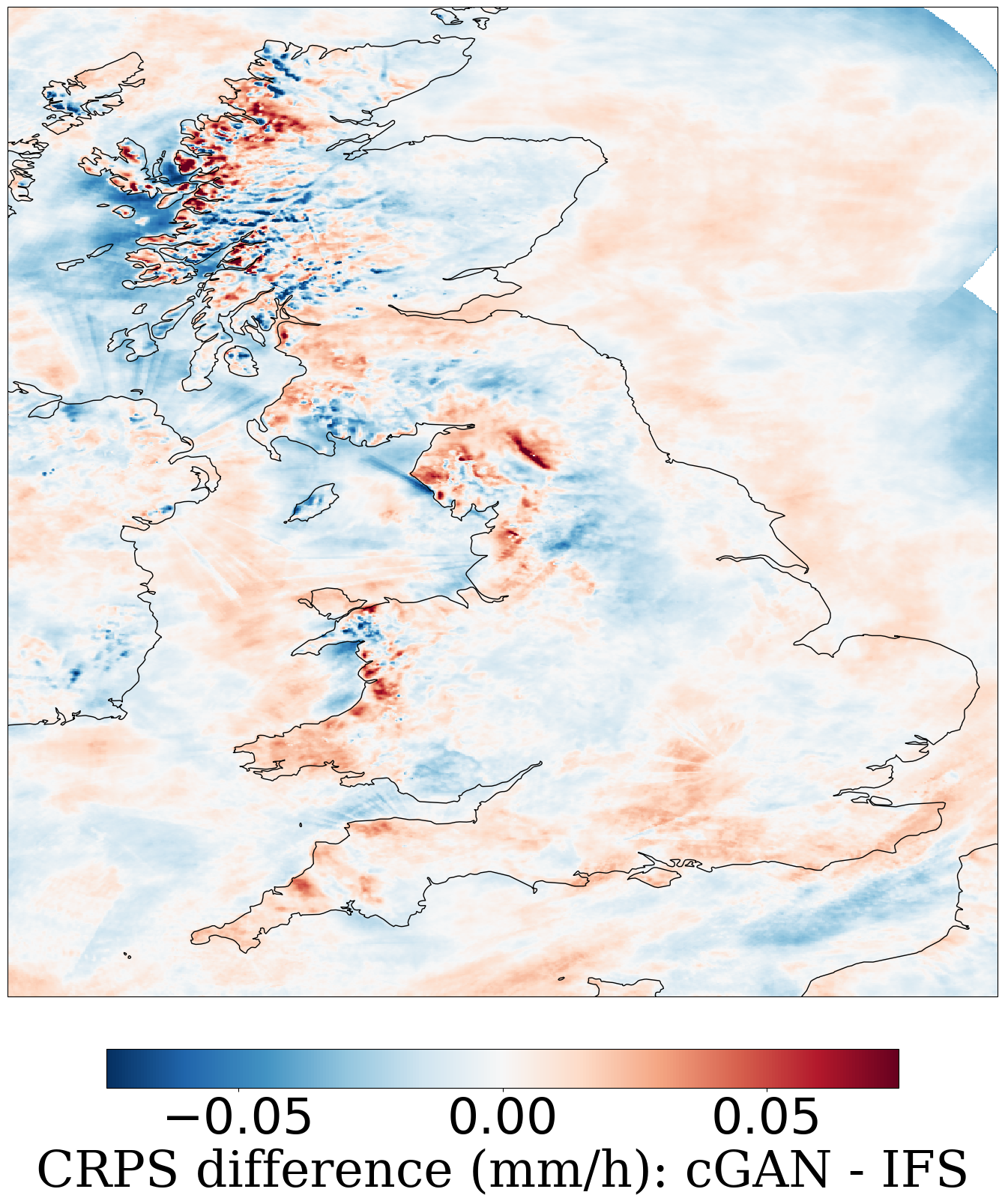} &
  \vspace{0pt} \includegraphics[width=0.29\textwidth]{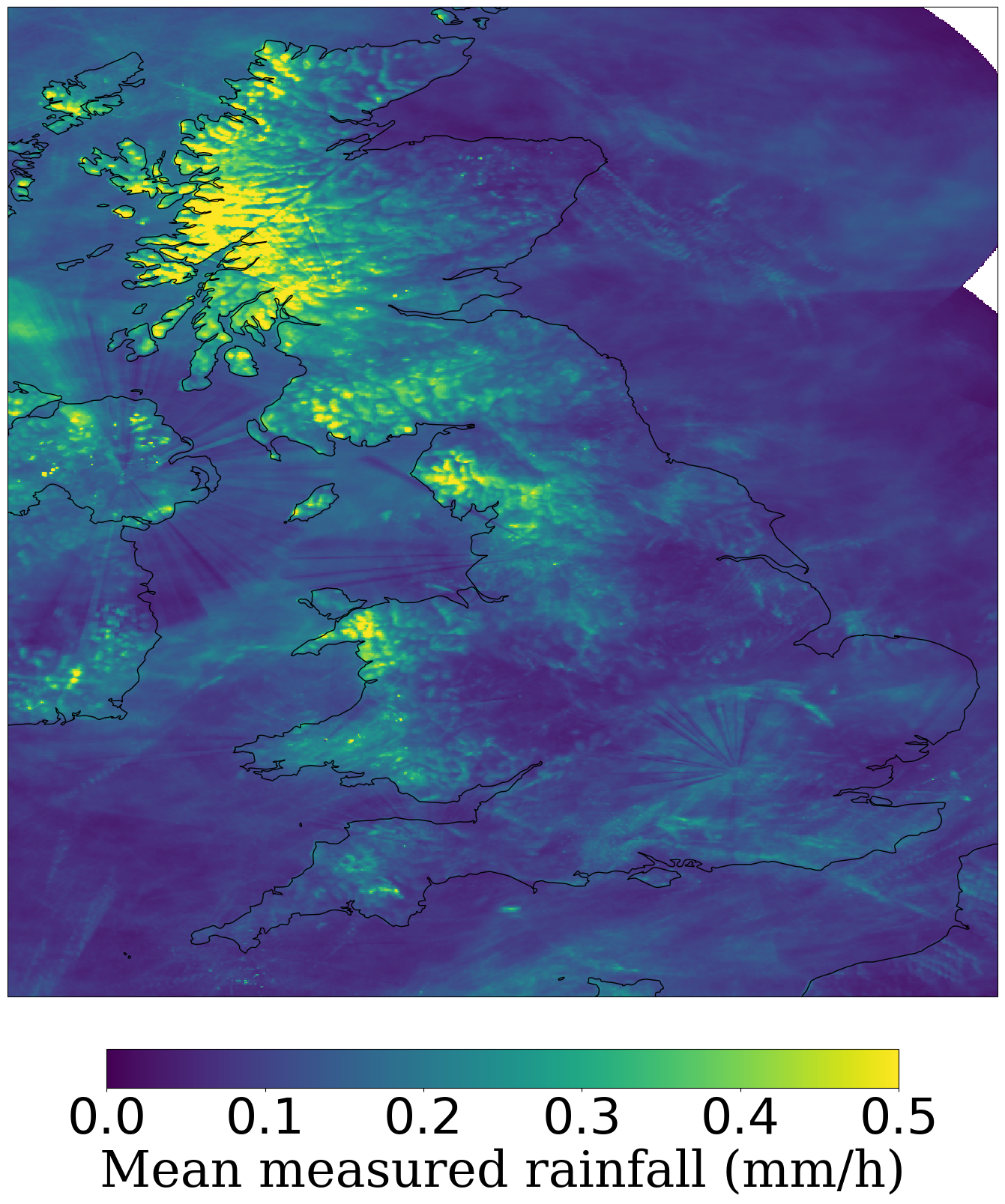} \\
\end{tabular}
\end{center}
\caption{{\bf Left:} CRPS (lower is better) at each grid point of the 50 member cGAN ensemble trained in each region separately. {\bf Middle:} Difference between the CRPS of cGAN and IFS. Blue indicates that cGAN has the lower (better) CRPS, red indicates that IFS has the lower CRPS. {\bf Right:} Average rainfall in each region. {\bf Top to bottom}: Pacific north-west, Great plains, Atlantic coast, UK.}
\label{CRPS:fig}
\end{figure}

The CRPS reported in table \ref{CRPS:table} is the mean of the CRPS computed at each $\sim 1$ km grid cell, which is plotted in the left column of figure \ref{CRPS:fig}. Comparison to the average rainfall (right column) indicates that the CRPS is dominated by where it is usually rains. The performance of cGAN relative the the IFS ensemble is plotted in the central column. In the Pacific north-west region (top, centred on Portland) cGAN has particular problems on the coast, near the edge of the domain. There are no ocean grid points represented at all and this might help explain the problem. However, in the hilly north of Scotland in the UK region (bottom) a similar problem occurs. This is somewhat offset by cGAN having lower CRPS scores over the neighbouring ocean, indicating potential regions for improvement in IFS. Other than that, cGAN is better in some regions and the IFS ensemble in others with no clear pattern. The particularly high rainfall region off the Atlantic coast (third from top, centred on Washington DC), is represented better by IFS and has a large contribution to the area mean CRPS, but the adjacent region of ocean is represented better by cGAN. Further investigation is required to quantify if this is due to a small number of weather events.

\begin{figure}
\begin{center}
\setlength\extrarowheight{-20pt}
\addtolength{\tabcolsep}{-7pt}
\begin{tabular}{p{0.3\textwidth} p{0.3\textwidth} p{0.4\textwidth}}
  \vspace{0pt} \includegraphics[width=0.29\textwidth]{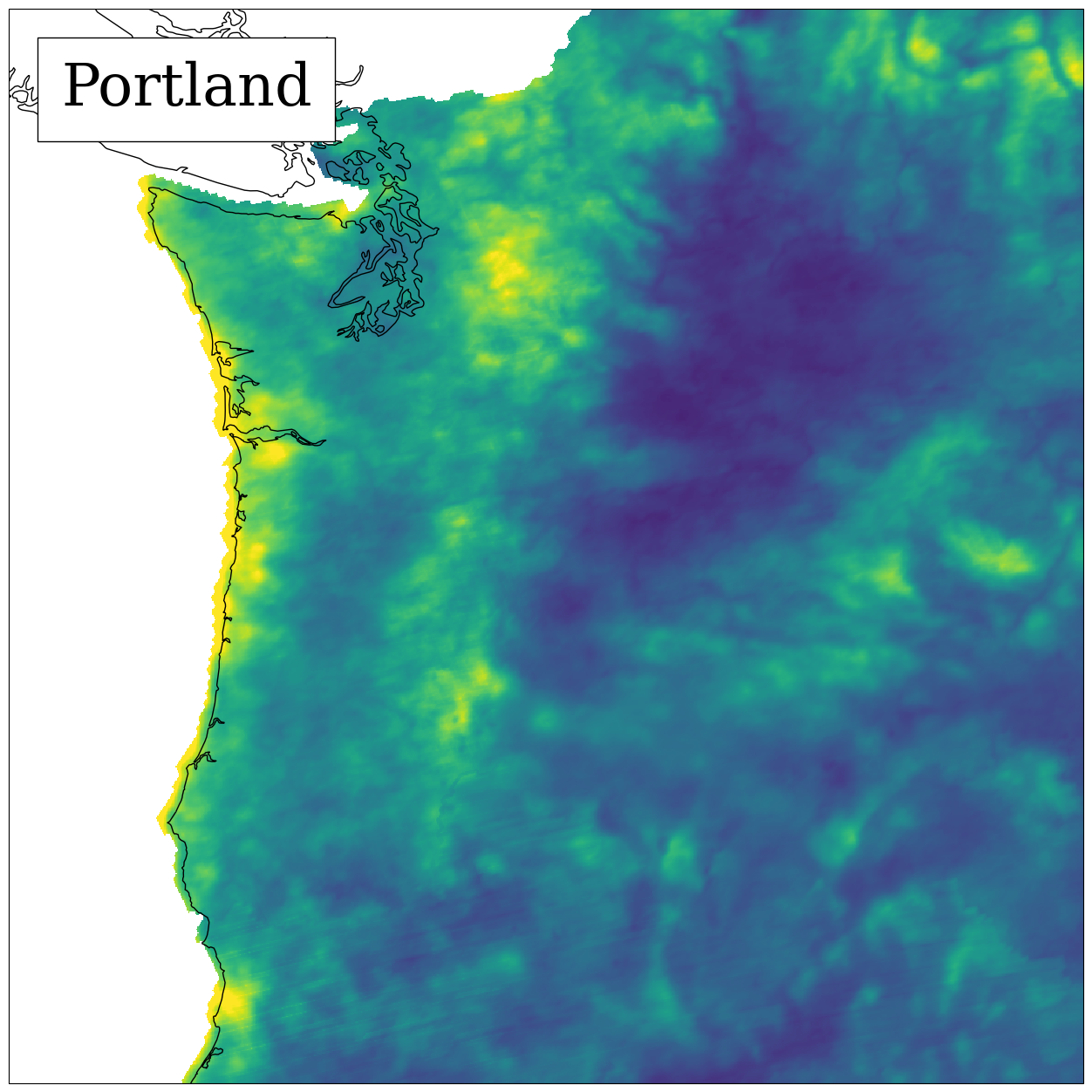} &
  \vspace{0pt} \includegraphics[width=0.29\textwidth]{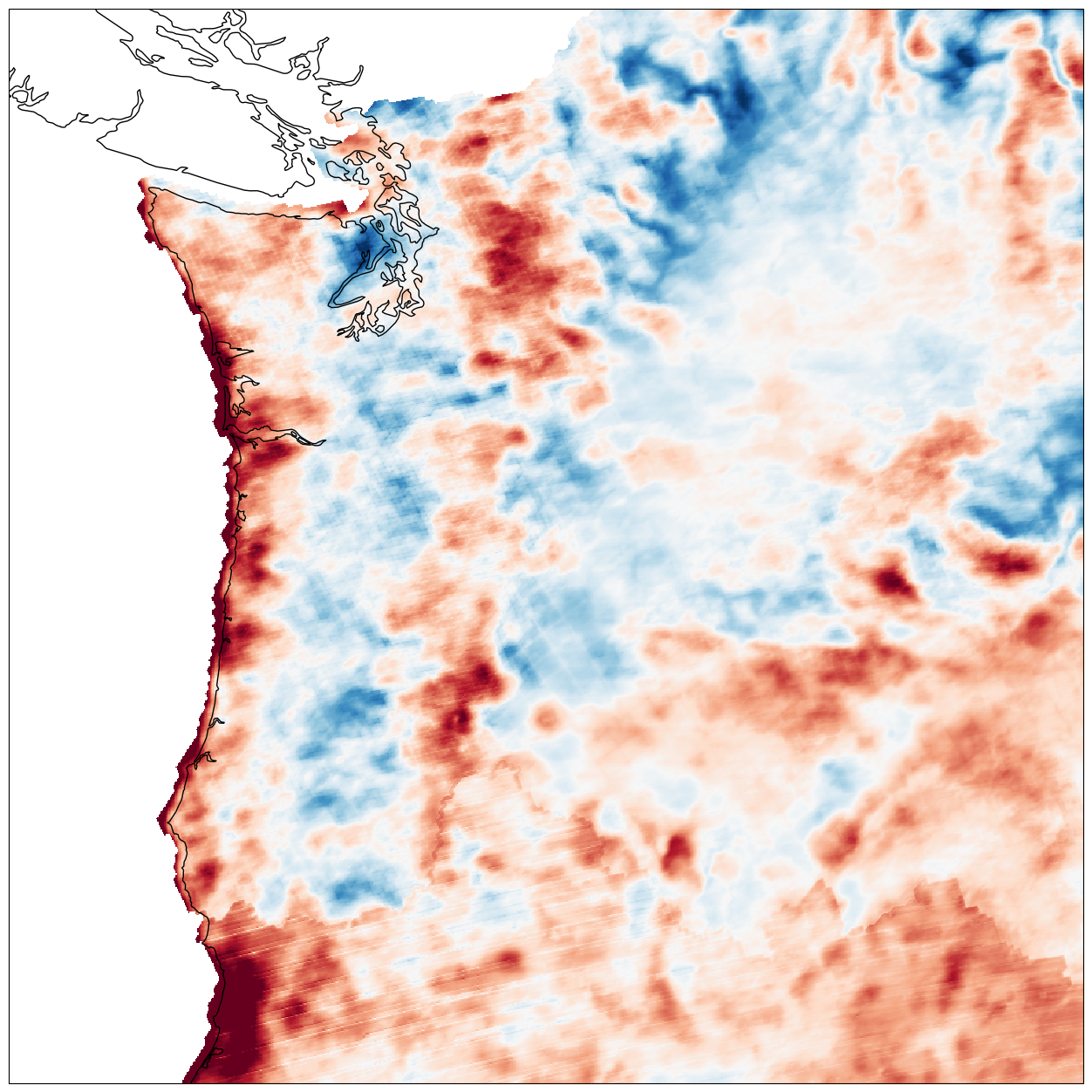} &
  \vspace{0pt} \includegraphics[width=0.405\textwidth]{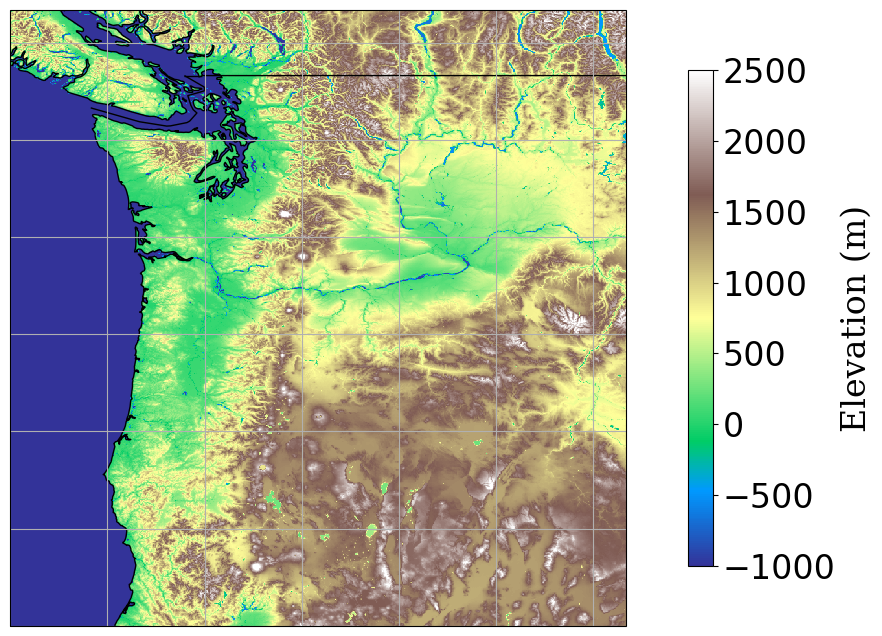} \\
  \vspace{0pt} \includegraphics[width=0.29\textwidth]{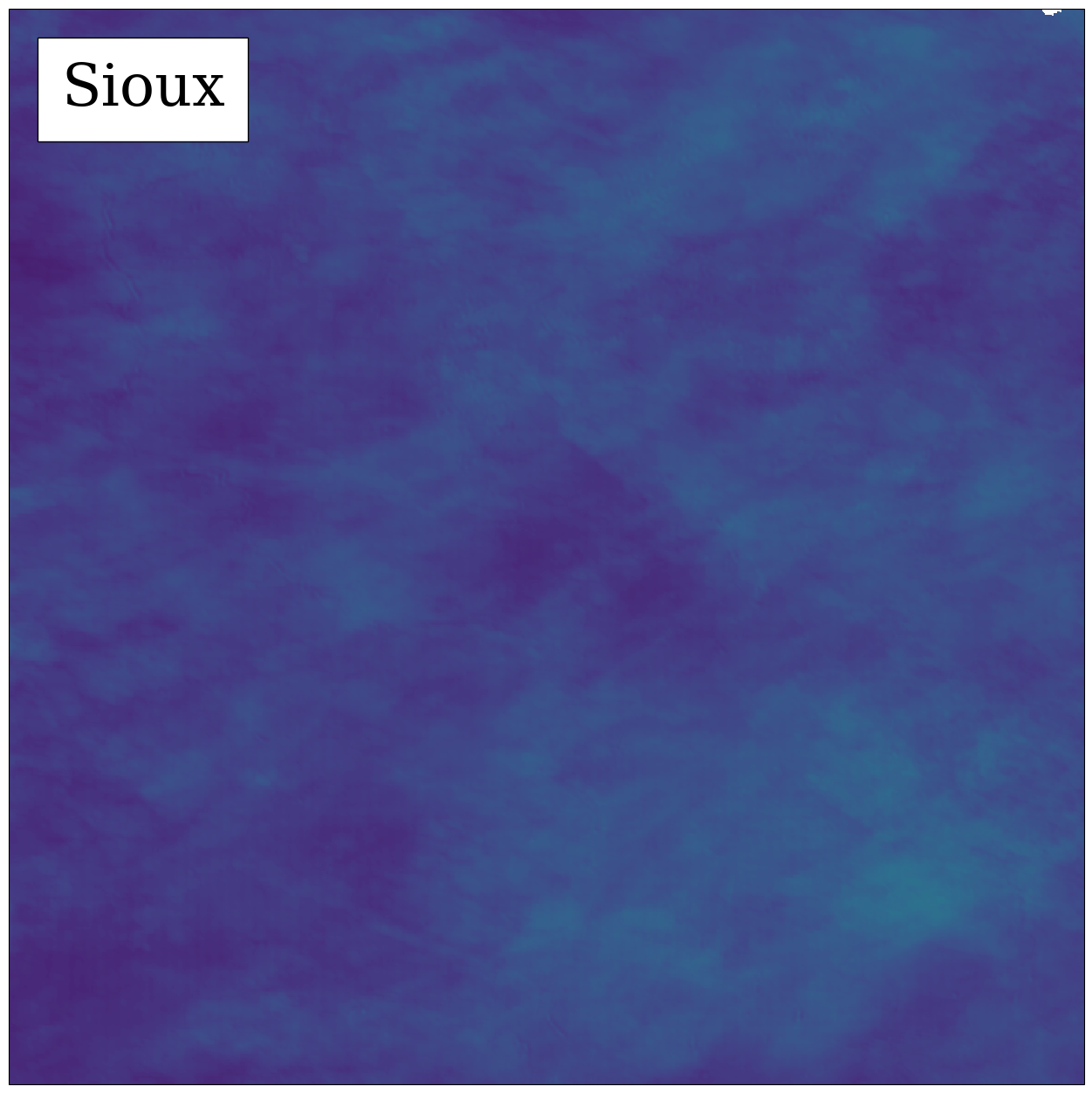} &
  \vspace{0pt} \includegraphics[width=0.29\textwidth]{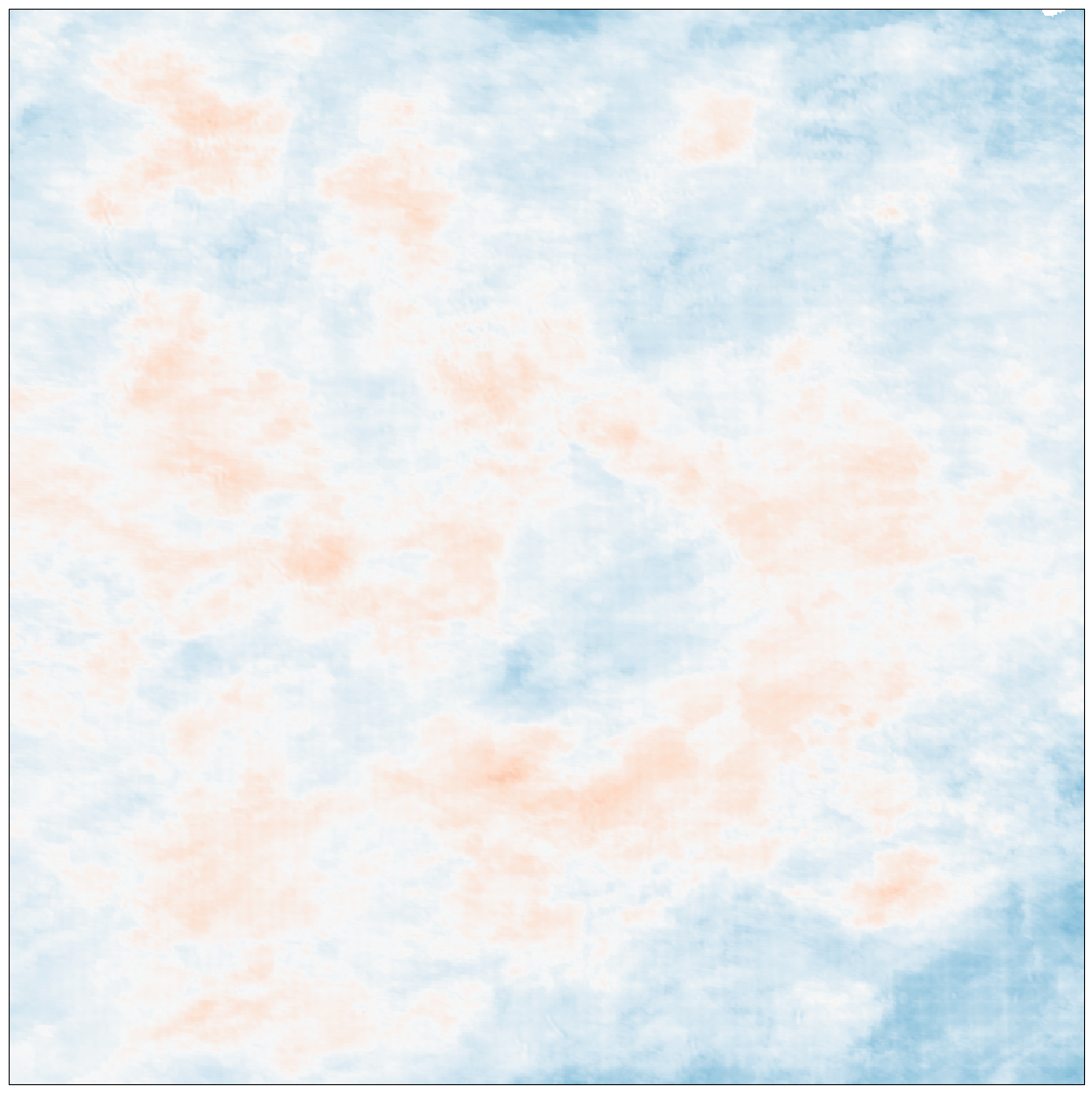} &
  \vspace{0pt} \includegraphics[width=0.395\textwidth]{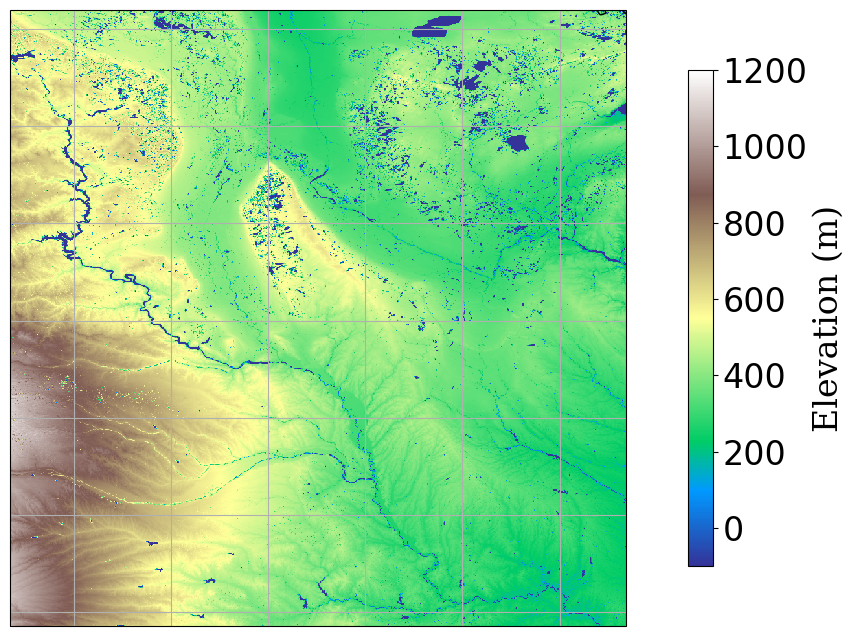} \\
  \vspace{0pt} \includegraphics[width=0.29\textwidth]{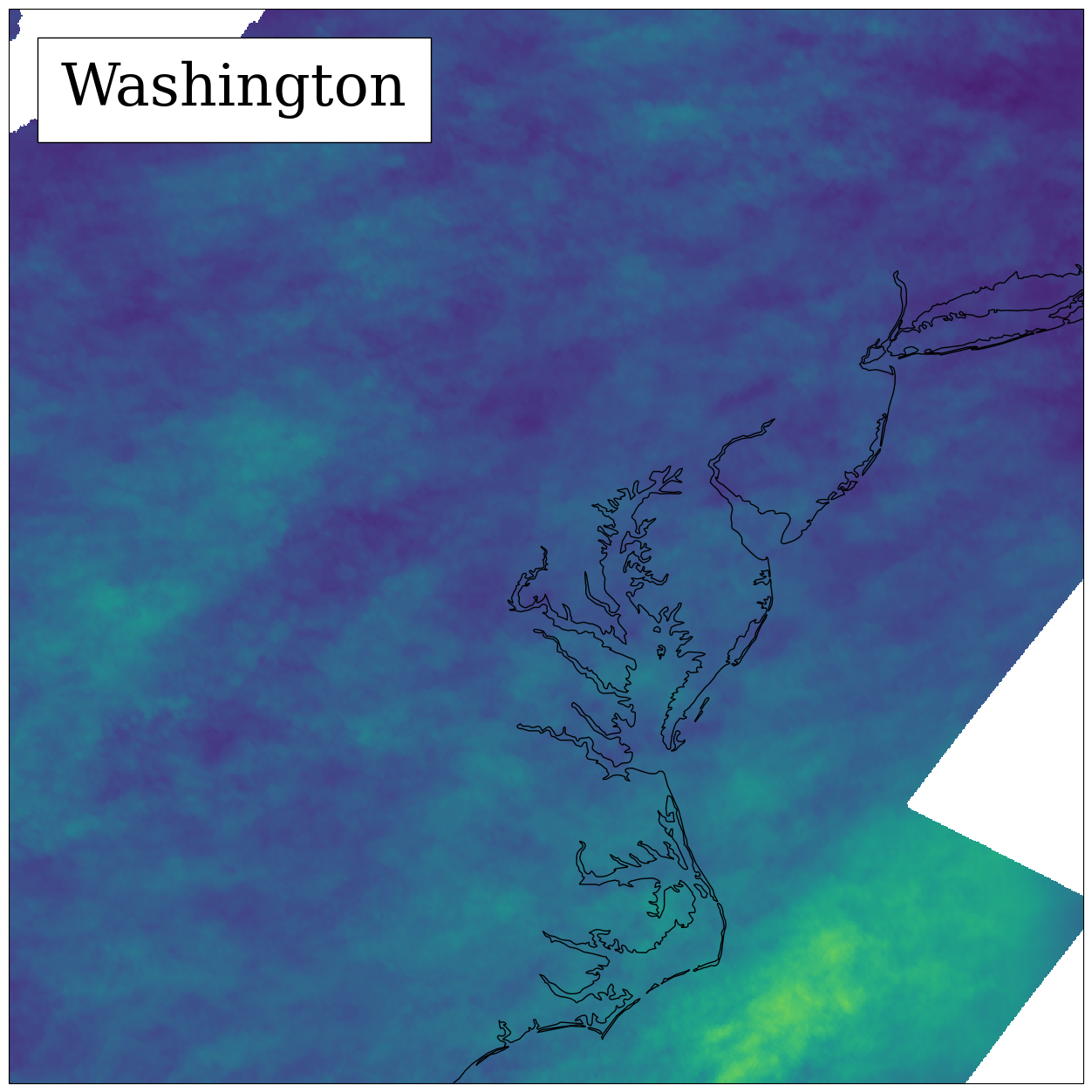} &
  \vspace{0pt} \includegraphics[width=0.29\textwidth]{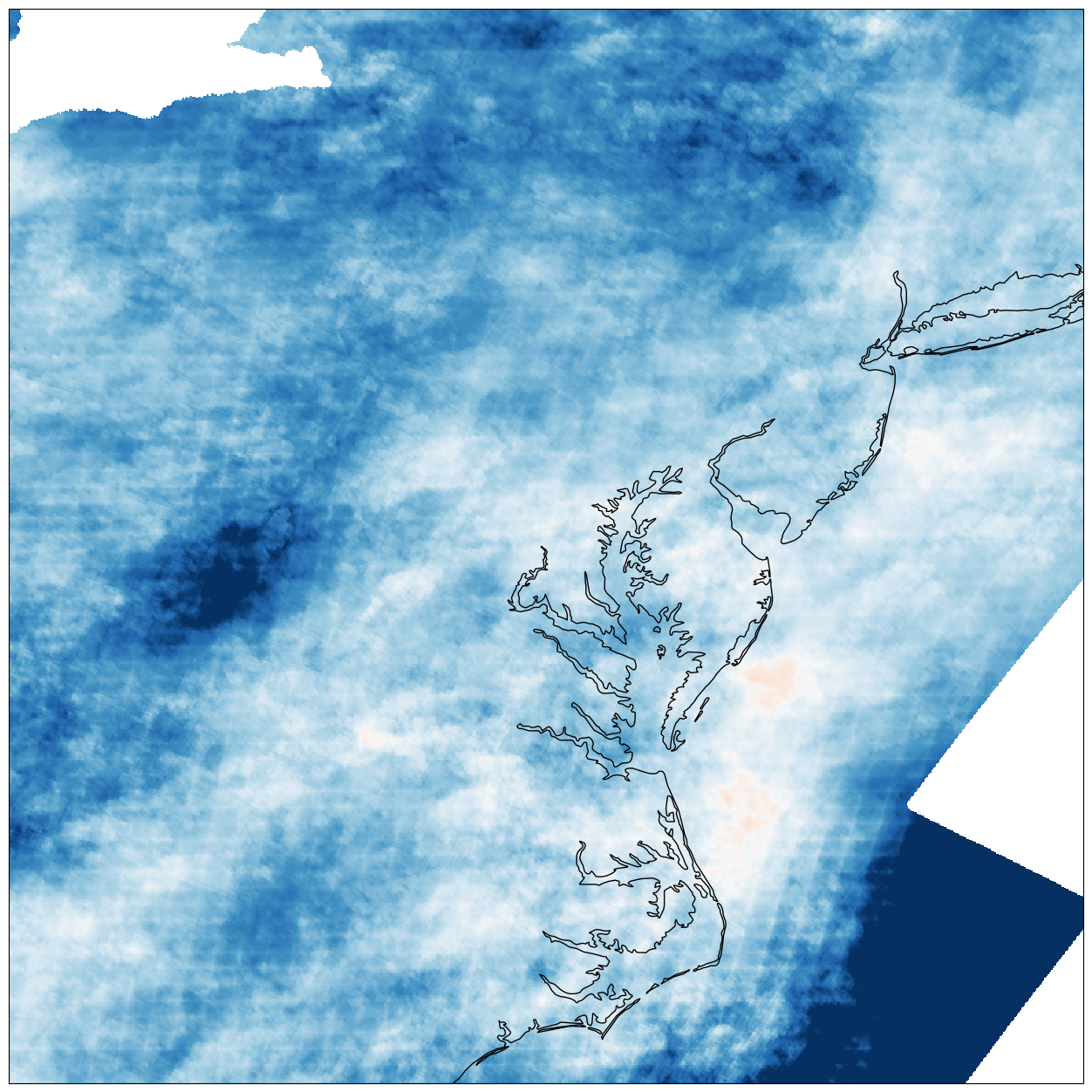} &
  \vspace{0pt} \includegraphics[width=0.405\textwidth]{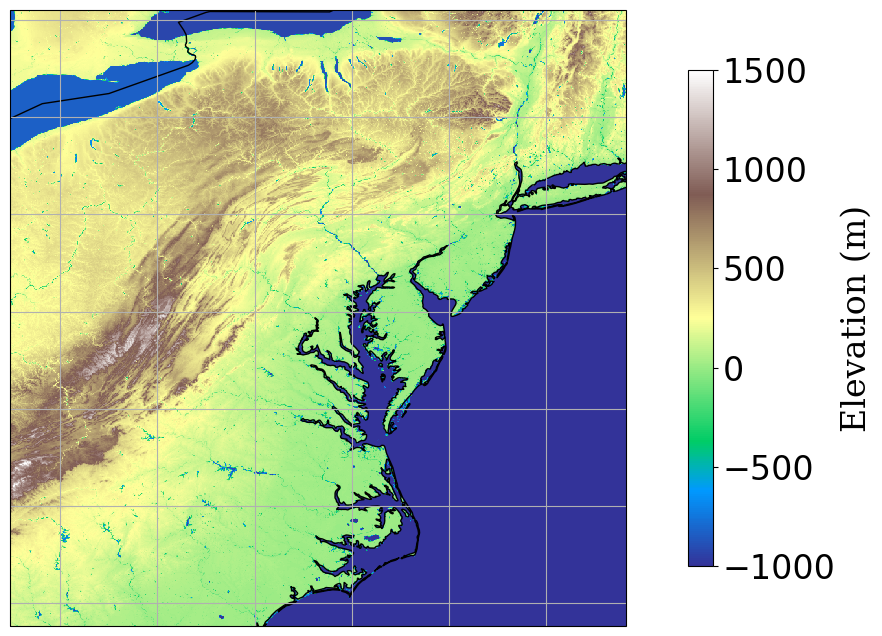} \\
  \vspace{0pt} \includegraphics[width=0.29\textwidth]{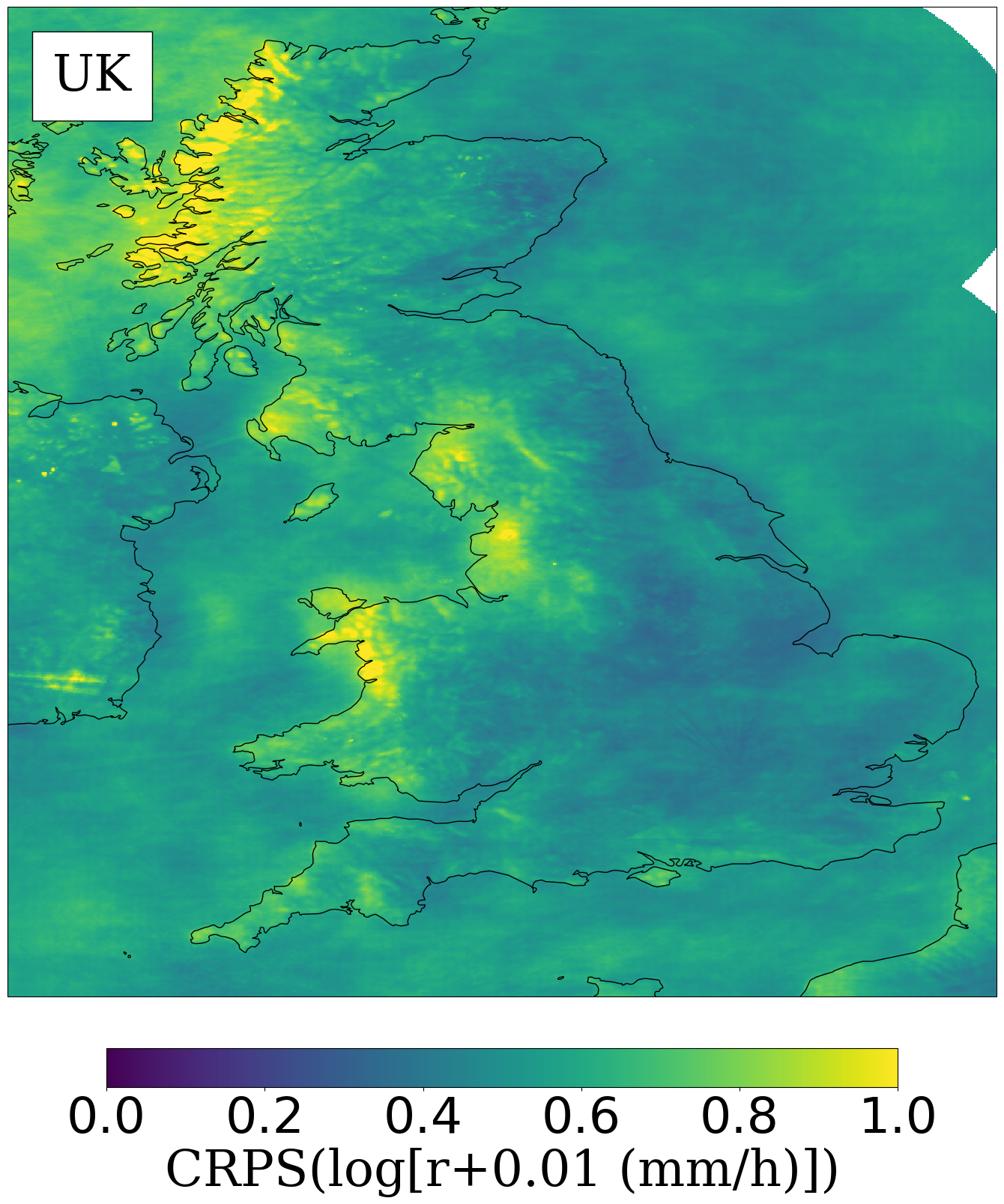} &
  \vspace{0pt} \includegraphics[width=0.29\textwidth]{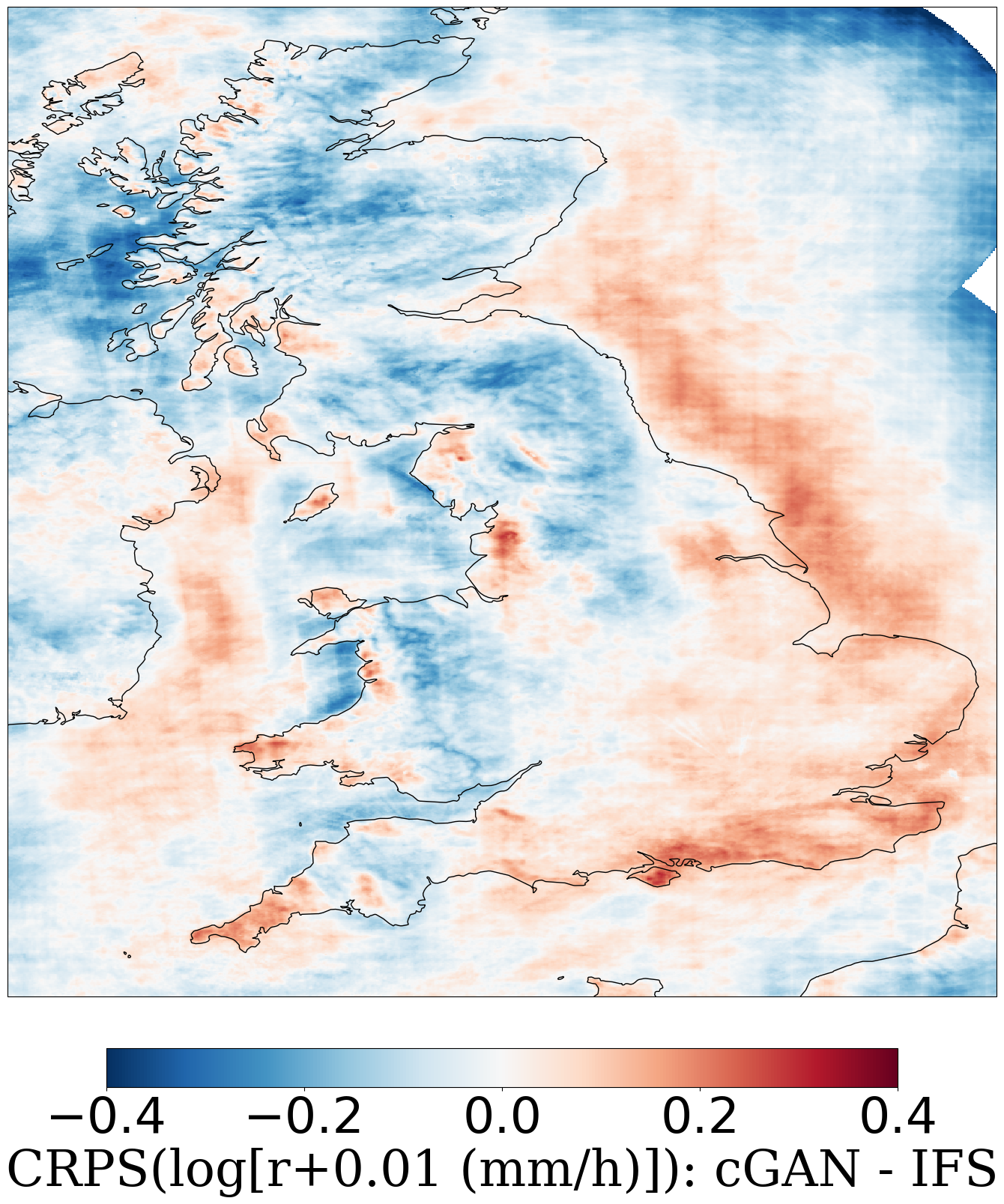} &
  \vspace{0pt} \includegraphics[width=0.395\textwidth]{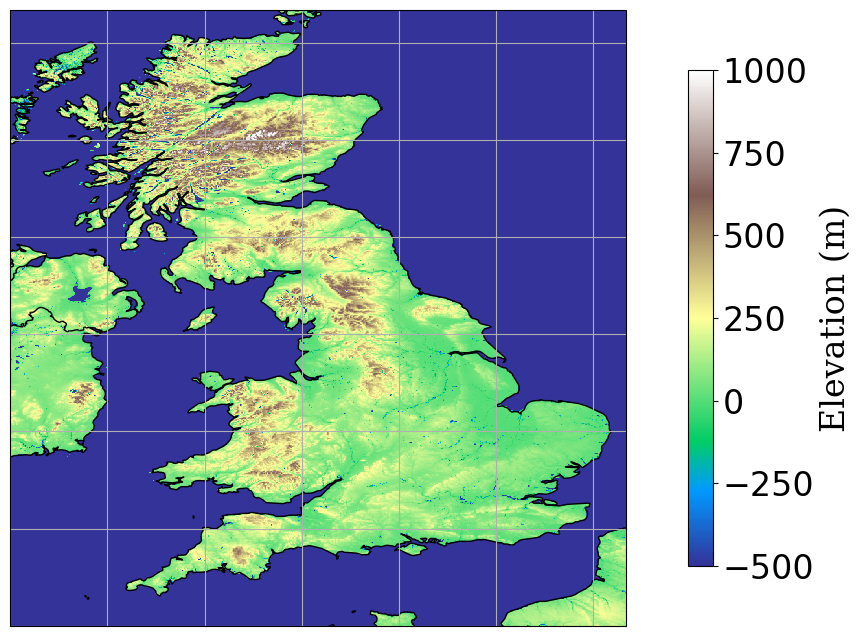} \\
\end{tabular}
\end{center}
\caption{{\bf Left:} CRPS(log($r$+0.01)) at each grid point of the 50 member cGAN ensemble trained in each region separately. $r$ represents rainfall in mm/h. {\bf Middle:} Difference between the CRPS(log(r+0.01)) of cGAN and IFS. Blue indicates that cGAN has the lower (better) CRPS(log(r+0.01)), red indicates that IFS has the lower CRPS(log(r+0.01)). {\bf Right:} Elevation in each region. {\bf Top to bottom}: Pacific north-west, Great plains, Atlantic coast, UK.}
\label{CRPS_log:fig}
\end{figure}

Rainfall has a particularly extreme distribution, with large regions of no rainfall at all and a high number of heavy rainfall events in the tail of the distribution. As shown in figure \ref{CRPS:fig}, the CRPS is dominated by the high rainfall regions. The distribution of the logarithm of rainfall, where it occurs, is far less extreme. For example, if the tail of the rainfall distribution follows a power law, then the tail of the distribution of logarithm of rainfall will decay exponentially. In order to examine the quality of the rainfall distribution away from its extremes, we plot the CRPS of $\log \left(r + 0.01 \text{mm/h}\right)$ where $r$ stands for rainfall, figure \ref{CRPS_log:fig}. The reason for adding 0.01 mm/h is to avoid having to deal with periods of zero rainfall while spreading the large number of low rainfall events across the distribution. Comparing the first columns of figures \ref{CRPS:fig} and \ref{CRPS_log:fig} shows that the spatial CRPS($\log(r+0.01)$) field is much smoother in space and is not entirely dominated by the rainfall quantity. In particular the uniformity in the Sioux region suggests much less of a dependence upon individual weather events. Comparison of the Portland region to the elevation (third column of figure \ref{CRPS_log:fig}) shows the dependence of the CRPS($\log(r+0.01)$) upon elevation, which is also somewhat related to rainfall quantity. However, the elevation in the Sioux and Washington regions appears to play much less of a role. In the UK, the CRPS($\log(r+0.01)$) still has a similar pattern to the rainfall quantity.

Comparison of CRPS($\log(r+0.01)$) between cGAN and the IFS ensemble is plotted in the middle column of figure \ref{CRPS_log:fig}. In the Portland region cGAN appears once again to have difficulty along the coastline and at high altitude, whilst getting improved results at lower elevations. In the region of Sioux city, cGAN and the IFS ensemble are pretty indistinguishable, while in the region centred on Washington DC cGAN has improved the CRPS($\log(r+0.01)$) almost everywhere. In both of these regions there seems to be little dependence of CRPS($\log(r+0.01)$) upon elevation. Over the UK, cGAN does poorly in some regions of the west coast, as with the CRPS, and IFS is better over the east coast and south east England. Overall the averages of these images correspond well to the respective averages in table \ref{CRPS:table}.

The root-mean-squared-error of the ensemble mean for each model is presented in table \ref{RMSEEM:table}. The numbers here tell much the same story as the CRPS scores in table \ref{CRPS:table}, indicating that this aspect of the rainfall distribution is similarly represented.

\begin{table}
\begin{center}
\begin{tabular}{l|cccc}
\toprule
& & Data & &  \\
Model        & Portland &   Sioux  & Washington &    UK \\
\midrule
cGAN Portland  &  {\bf 0.341}  &  0.614  & 1.074 &   0.698 \\
cGAN Sioux     &  0.379  &  0.601 &  0.878  &  0.518 \\
cGAN Washington & 0.362  &  {\bf 0.581} &  0.828  &  0.447 \\
cGAN UK       &   0.416  &  0.622 &  0.879 &   {\bf 0.429} \\
cGAN UK HRES & 0.405 & 0.593 & 0.853 & 0.459 \\
cGAN All regions & 0.349 & 0.582 & {\bf 0.816} & 0.437 \\
IFS Ensemble & 0.385 & 0.681 & 0.921 & 0.453 \\
ERA5 (RMSE) & 0.349 & 0.625 & 0.881 & 0.472 \\
Zeros (RMSE) & 0.483 & 0.634 & 0.941 & 0.509 \\
\bottomrule
\end{tabular}
\caption{Root-Mean-Squared Error of the Ensemble Mean (RMSEEM, lower is better) of cGAN trained on and generating an ensemble of rainfall forecasts using ERA5 data in all cases except cGAN UK HRES that was trained on and uses HRES data for forecasts.}
\label{RMSEEM:table}
\end{center}
\end{table}

\subsection{Spatial variability}

The Radially Averaged Log Spectral Distance (RALSD) for each of the models is presented in table \ref{RALSD:table}. With the exception of the Sioux model, each model appears to optimise these scores for its own region. In contrast to the CRPS, although the Washington model is good in the USA, it does less well over the UK. And vice-versa for the UK and UK HRES models. cGAN trained on all regions does not have particularly impressive scores by this metric and does not reproduce the power spectrum as well as the local models. A similar story applies to the corresponding variogram scores in table \ref{variogram:table} although the cGAN UK HRES and cGAN all region models appear to do much better.

For the deterministic forecasts IFS HRES scored similarly to IFS ensemble member 2 and surprisingly worse than the lower resolution ERA5. The cGAN HRES was the best model over the UK however it performed badly in the USA, considerably worse than IFS ensemble member 2. It is not clear why this should be the case. Further investigation is required to determine if it was due to potential outlier ensemble members or some other cause. We have pretty much the same story for the variogram scores of $\log(r+0.01)$ (not shown), with the exception that the IFS ensemble over the UK is relegated to score below the cGAN UK models (which did better) and has a similar score to the cGAN all region model.

\begin{table}
\begin{center}
\begin{tabular}{l|cccc}
\toprule
& & Data & &  \\
Model        & Portland &   Sioux  & Washington &    UK \\
\midrule
cGAN Portland   & 1.607 & 5.524 & 2.333 & 4.792 \\
cGAN Sioux       & {\bf 1.144} & 6.886 & {\bf 1.026} & 3.935 \\
cGAN Washington & 4.659 & 3.508 & 1.427 & 12.047 \\
cGAN UK         & 4.665 & 10.357 & 7.066 & 1.473 \\
cGAN UK HRES & 11.321 & 5.595 & 4.873 & {\bf 0.792} \\
cGAN All regions & 4.202 & {\bf 1.377} & 1.528 & 5.223 \\
IFS Ensemble & 6.896 & 7.621 & 8.164 & 15.042 \\
IFS Ens. member 2 & 5.407 & 6.173 & 7.285 & 14.836 \\
IFS HRES & 3.445 & 4.775 & 5.672 & 13.373 \\
ERA5 & 5.860 & 9.012 & 7.775 & 15.031 \\
\bottomrule
\end{tabular}
\caption{Radially Averaged Log Spectral Distance (RALSD, lower is better) of cGAN and the linearly interpolated IFS as in table \ref{RMSEEM:table}. For more insight see figure \ref{RAPSD:fig}.}
\label{RALSD:table}
\end{center}
\end{table}

Like the CRPS, the variogram score in table \ref{variogram:table} is the average of the variogram scores at each grid location. Plotting these maps (not shown) indicates that like the CRPS, the variogram score is high in the regions of high rainfall. There were also some isolated places, for example in the south east of Ireland, where the variogram score was high. Inspection of the NIMROD data in this region and the Stage IV data in others revealed a number of radar artefacts, indicating that the variogram score is sensitive to them.

\begin{table}
\begin{center}
\begin{tabular}{l|cccc}
\toprule
& & Data & &  \\
Model        & Portland &   Sioux  & Washington &    UK \\
\midrule
cGAN Portland   & 1.479 & 2.377 & 5.791 & 4.928 \\
cGAN Sioux       & 1.511 & 2.283 & 4.313 & 3.119 \\
cGAN Washington & 1.360 & 1.805 & 3.762 & 2.618 \\
cGAN UK         & 2.027 & 3.318 & 4.675 & 2.358 \\
cGAN UK HRES & 1.588 & {\bf 1.761} & 3.602 & 2.197 \\
cGAN All regions & {\bf 1.255} & 1.921 & {\bf 3.396} & 2.321 \\
IFS Ensemble & 2.665 & 2.752 & 5.772 & {\bf 2.046} \\
IFS Ens. member 2 & 1.714 & 2.240 & 4.343 & 2.673 \\
IFS HRES & 1.786 & 2.168 & 4.417 & 2.579 \\
ERA5 & 1.581 & 1.979 & 3.883 & 2.615 \\
\bottomrule
\end{tabular}
\caption{Variogram scores (lower is better) of cGAN and the linearly interpolated IFS as in table \ref{RMSEEM:table}.}
\label{variogram:table}
\end{center}
\end{table}

\begin{figure}
\begin{center}
\includegraphics[width=0.49\textwidth]{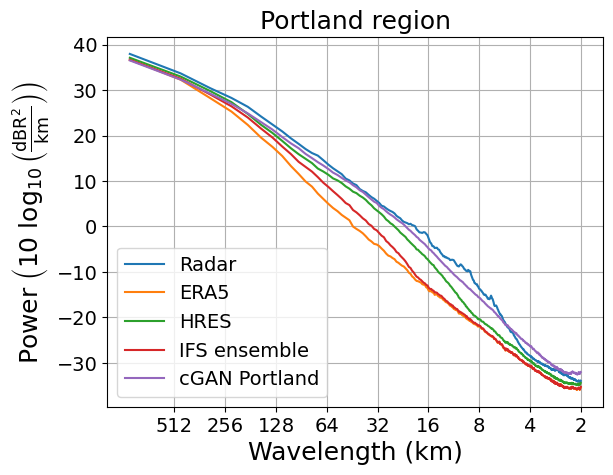}
\includegraphics[width=0.49\textwidth]{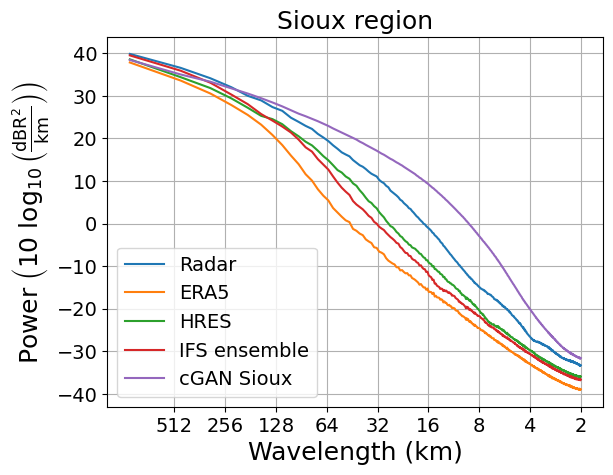}
\includegraphics[width=0.49\textwidth]{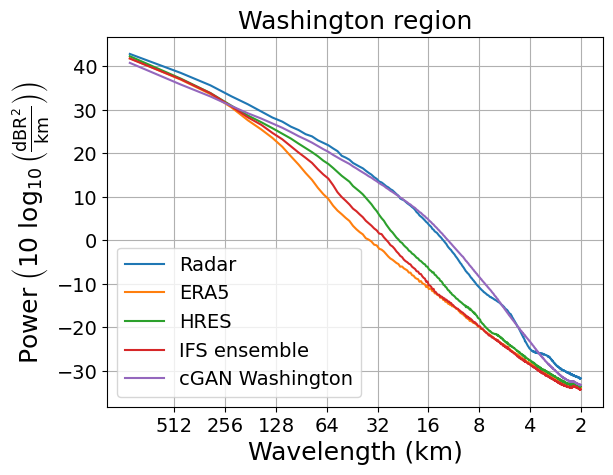}
\includegraphics[width=0.49\textwidth]{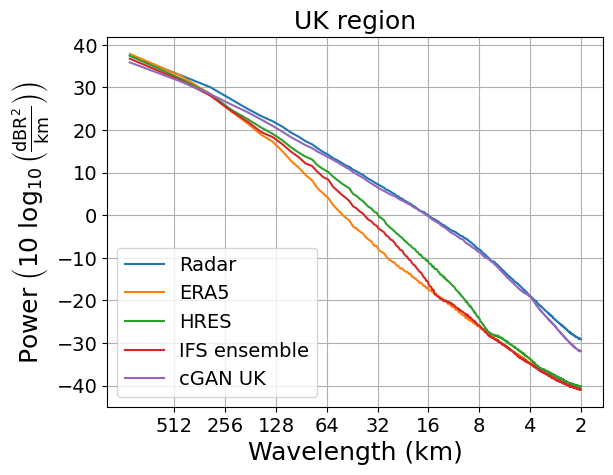}
\end{center}
\caption{The radially averaged power spectra for models over each of the four regions. In contrast to \citeA{Harris22}, these curves are averages over all sample dates. They represent $\overline{P}_{\text{radar},i}$ and $\overline{P}_{\text{model},i}$ used to calculate the RALSD in equation \ref{RALSD:eqn}. Compare to table \ref{RALSD:table}.}
\label{RAPSD:fig}
\end{figure}

The radially averaged power spectra used to compute the RALSD are plotted in figure \ref{RAPSD:fig}. At low frequencies the curves are proportionally close together on the log scale. For all regions, as the frequency increases the interpolated IFS forecast models fall below the measurement data in order of their resolution. This classical resolution dependent behaviour can be understood to reflect the ability of a fluid model to represent different scales. Smoothing the rainfall data achieves similar, though not identical, curves. Starting from the ERA5 model, in each region cGAN has managed to generate the high frequency variability necessary to correct these curves. Although, the power spectrum in the Sioux city region was not corrected as well as in the other regions. Note that otherwise ``good'' models, such as cGAN HRES and cGAN all regions performed poorly with this metric when compared to the local models.

\subsection{Orography resolution}

In these experiments cGAN over the USA was trained using $\sim 1$km orography and land-sea mask derived from the GMTED2010 and ESA WorldCover 2020 data sets. For the UK we, and \citeA{Harris22}, use the lower resolution orography and land-sea mask field used with the IFS. To see the impact, we also trained the ERA5 cGAN UK model with different orography fields. The results, summarised in table \ref{orography:table}, indicate that using the low resolution IFS ensemble field or GMTED2010 has only a marginal if any impact upon the CRPS, RMSEEM and Variogram score with respect to using the higher resolution field. However the RALSD was degraded. This degradation might be within the random variation seen between training runs though, compare to table \ref{retrain:table} below.

\begin{table}
\begin{center}
\begin{tabular}{l|cccc}
\toprule
Orography      & CRPS  & RMSEEM & RALSD & Variogram \\
\midrule
GMTED2010  & 0.0970 & 0.459 & 1.736 & 2.357 \\
IFS high resolution & 0.0972 & 0.434 & 1.473 & 2.358 \\
IFS operational & 0.0970 & 0.432 & 2.197 & 2.336 \\
\bottomrule
\end{tabular}
\caption{cGAN trained over the UK on ERA5 with orography and the land-sea mask from three different sources. The highest resolution orography is GMTED2010 and the lowest is that from the operational IFS.}
\label{orography:table}
\end{center}
\end{table}

\subsection{Model variation}

To understand the uncertainty in the model training, which is a somewhat random process, the models in each region are trained again from scratch. They are then evaluated again using the same dates and times used in the first training-evaluation. The resulting scores (table \ref{retrain:table}) give some indication of the range of uncertainty in the training procedure. Using the same dates (within each region) reduces the variability in the scores making them easier to compare. The CRPS, RMSEEM and Variogram scores appear quite robust over the two training runs. The same cannot be said for the RALSD. This is despite the RALSD being quite robust within independent sample dates using a single model.

To measure the true uncertainty in the scores, the evaluation could be computed multiple times using random dates within the test year. However we found that this introduces an unacceptably large weather induced variability. The scores also depend on the precise selection of region and are only used to compare models, so reducing variation is the priority.

Training models is the most computationally intensive part of this work. The cost of additional training runs to more accurately quantify the training uncertainty is currently prohibitive. Unfortunately, with only two data points on the training axis, there is nothing to be gained by employing any statistical techniques.

\begin{table}
\begin{center}
\begin{tabular}{l|cccccc}
\toprule
Region      & CRPS  & RMSEEM & RALSD & RALSD & RALSD & Variogram \\
& & & & first 128 & last 128 & \\
\midrule
Portland 1 & 0.068 & 0.341 & 1.607 & 1.800 & 1.441 & 1.479 \\
Portland 2 & 0.064 & 0.334 & 2.055 & 2.363 & 1.701 & 1.300 \\
\midrule
Sioux 1 & 0.059 & 0.601 & 6.886 & 5.701 & 7.951 & 2.283 \\
Sioux 2 & 0.062 & 0.585 & 1.838 & 1.642 & 2.109 & 2.333 \\
\midrule
Washington 1 & 0.112 & 0.828 & 1.427 & 1.165 & 1.715 & 3.762 \\
Washington 2 & 0.112 & 0.831 & 3.206 & 2.478 & 4.100 & 3.454 \\
\midrule
UK 1 & 0.096 & 0.429 & 1.473 & 1.042 & 2.011 & 2.358 \\
UK 2 & 0.094 & 0.419 & 1.021 & 1.132 & 1.018 & 2.271 \\
\bottomrule
\end{tabular}
\label{retrain:table}
\caption{Variation in scores in each regions with different training runs of each model. The evaluation of the scores is performed using the same dates and times in each region. For the RALSD additional scores were averaged over the first and last 128 sample dates as well as all 256, the default used here.}
\end{center}
\end{table}

\section{Discussion and conclusions}

Application of the cGAN model developed in \citeA{Harris22} to 3 additional regions over the USA leads to broadly similar results to those reported for the UK. cGAN has been shown here to be capable of post-processing low resolution ERA5 data into an ensemble competitive with the IFS ensemble and IFS HRES forecast at short lead times as measured by the CRPS. In addition cGAN was able to ``downscale'' the ERA5 rainfall data to $\sim 1$ km resolution and correct the high spatial frequency variability towards that of the measured rainfall, outperforming the IFS ensemble and IFS HRES forecast over the USA.

In general, models trained upon a particular region did well at predicting the rainfall in that region, but less well elsewhere. This could be for example because of dynamics or conditions specific to each region, or possibly a lack of training data. Over the two to three years of training data, there were long periods where large areas remained dry. Particularly in the regions centred on Portland and Sioux city where the local cGAN models performed less well. In an attempt to account for this lack of data another cGAN model was trained using all of the training data from all four regions. Unlike the other models, this model might find it harder to take advantage of dynamics specific to the local region. In addition, the character of the radar data is different in each region: $\sim 4$ km for Stage IV and $\sim 1$ km for NIMROD, with each region in the USA and UK using different processing algorithms and calibration. Although the radar hardware is similar. Despite these disadvantages, the cGAN model using training data from all regions equaled or outperformed the local models everywhere.

If it is the case that the ensembles produced by cGAN model are improved by simply using more training data, there are many rainfall radar stations around the world online now, in addition to the regions of the USA that were not considered here. However, quality and characteristics of the gridded products are not uniform and challenges remain. Africa for example has no available rainfall radar station data that we are aware of.

To further validate the methodology, the model using training data from all regions should be tested elsewhere where there is radar. For example Taiwan, where rainfall forecasts can be challenging. All four regions in this paper are in the extra-tropics, so it is not clear if cGAN is limited to the large scale rainfall patterns in the extra-tropics. Training and testing cGAN in tropical regions is difficult due to the lack of ground based radar data. We are currently investigating this direction in the horn of Africa region using satellite measurements of rainfall, which do not occur as often and are reported at a lower resolution compared to ground based radars. We are also investigating post-processing of forecasts into the medium range future and using information from the entire forecast ensemble rather than a single data set.

UK model evaluation suggests that using a better dynamical forecast model results in improved cGAN predictions, in both training and evaluation.

\appendix
\section{Reduced Numerical Precision}
\label{precision:sect}

The models were trained on an NVIDIA A100 accelerator, on which TensorFlow automatically employed the “TensorFloat-32” (TF-32) format for many internal calculations. This number format has the range of 32-bit numbers but the precision of 16-bit numbers and has the advantage of lower computational cost compared to full 32-bit computations. Negligible impact of this reduced precision in the UK model trained using IFS data was reported by \citeA{Harris22}.

\section{Fractions skill score (FSS) example}
\label{FSS:sect}

\begin{figure}
\begin{center}
\includegraphics[width=0.44\textwidth]{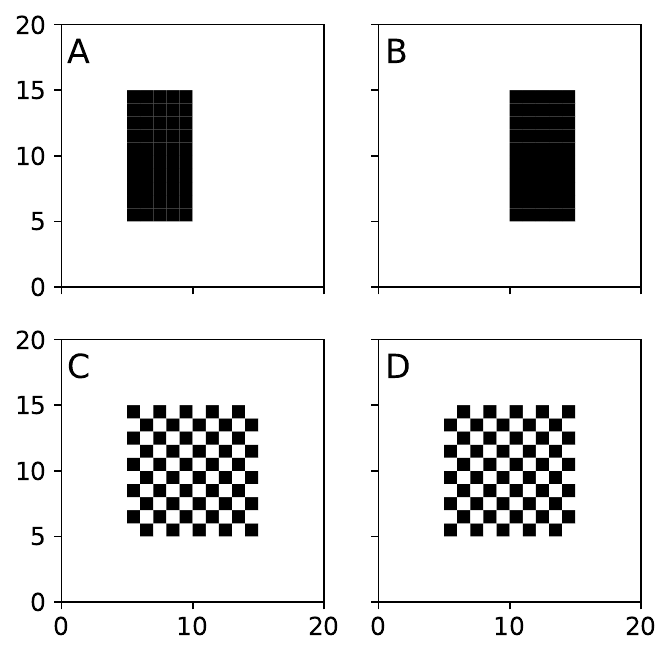}
\includegraphics[width=0.55\textwidth]{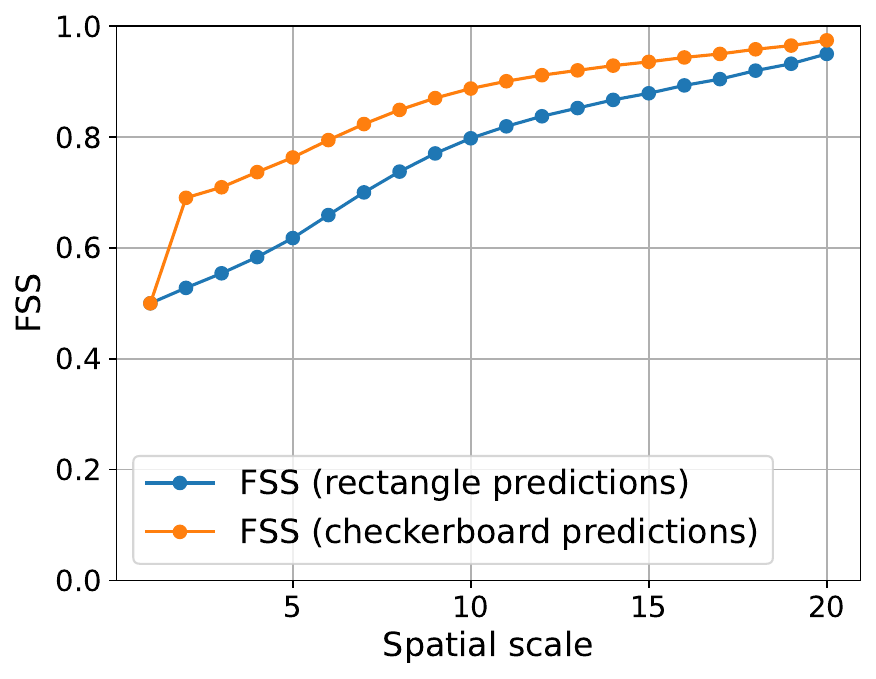}
\end{center}
\caption{{\bf Left}: Four example fields. A and B correspond to the truth 50\% of the time each and a ``good'' forecast of the truth, randomly A or B. C and D chosen randomly correspond to ``bad'' forecasts of A and B. {\bf Right}: The fractions skill score as a function of scale.
}
\label{FFSeg:fig}
\end{figure}

A perfect forecast has a FSS of 1, and a ``no skill'' forecast has a FSS of 0. Suppose we have a field, figure \ref{FFSeg:fig}, that looks like A 50\% of the time and B 50\% of the time. If we forecast our field by randomly choosing A and B with equal probability, then we get a lower (worse) fractions skill score than if we randomly forecast C and D instead. For this example the variogram score (lower is better) with unit weights and $p=0.5$ returns 75 for forecasting A and B and 100 for forecasting C and D.

%



%
%

\section*{Open Research Section}
The  code  for  the  GAN  and  VAE-GAN  models  used  in  this  paper  is  available  at  \url{https://doi.org/10.5281/zenodo.6922291}. A cleaned-up version of the code, with the same core functionality, is available at \url{https://github.com/ljharris23/public-downscaling-cgan}. We would recommend this for people looking to build on our work. Our code was adapted from Jussi Leinonen's GAN model, available at \url{https://github.com/jleinonen/downscaling-rnn-gan}. All experiments in this paper were performed within TensorFlow 2.7.0. The ECMWF forecast archive can be obtained through MARS; more details are available at \url{https://www.ecmwf.int/en/forecasts/access-forecasts/access-archive-datasets}.  MARS  accounts  for  academic  use  are  available  for  free,  subject  to  certain  conditions;  see  \url{https://www.ecmwf.int/en/forecasts/accessing-forecasts/licences-available}.  The  NIMROD radar data set can be obtained through CEDA; more details are available at \url{https://catalogue.ceda.ac.uk/uuid/27dd6ffba67f667a18c62de5c3456350}. A CEDA Archive account is required in order to access this data. The Stage IV data may be obtained at \url{https://data.eol.ucar.edu/dataset/21.093}.

\acknowledgments
This project has received funding from the European Research Council (ERC) under the European Union's Horizon 2020 research and innovation programme (Grant No 741112, ITHACA). MC gratefully acknowledge funding from the MAELSTROM EuroHPC-JU project (JU) under No 955513. The JU receives support from the European Union's Horizon research and innovation programme and United Kingdom, Germany, Italy, Luxembourg, Switzerland, and Norway.

%
%

\bibliography{references}

%
%
%
%
%

\end{document}